%% file: quant_receiver.tex
\documentclass[journal]{IEEEtran}
\ifCLASSINFOpdf
   \usepackage[pdftex]{graphicx}
\else
\fi
\usepackage[cmex10]{amsmath}
\usepackage{amssymb}
\usepackage{mathtools}
\usepackage{array}

\usepackage{algpseudocode,algorithm,algorithmicx}
\usepackage{float}
\usepackage{subfigure}
\usepackage{tabularx}
\usepackage{mathrsfs} 
\usepackage{psfrag}
\usepackage{tikz}
\usetikzlibrary{chains, arrows,shapes,positioning,arrows,decorations.markings,calc}
\usepackage{pgfplots}
\pgfplotsset{compat=newest}
\usepgfplotslibrary{groupplots}
\usepackage[hyphens]{url}
\usepackage{gensymb}
\usepackage{xcolor}
\usepackage{todonotes} 

\definecolor{GTOBMIX1}{rgb}{0,0.80,0.20}
\definecolor{GTOBMIX2}{rgb}{0,0.60,0.40}
\definecolor{GTOBMIX3}{rgb}{0,0.40,0.60}
\definecolor{GTOBMIX4}{rgb}{0,0.20,0.80}

\definecolor{BTORMIX1}{rgb}{0.125, 0,0.875}
\definecolor{BTORMIX2}{rgb}{0.25, 0,0.75}
\definecolor{BTORMIX3}{rgb}{0.375,0,0.625}
\definecolor{BTORMIX4}{rgb}{0.50,0,0.50}
\definecolor{BTORMIX5}{rgb}{0.625,0,0.375}
\definecolor{BTORMIX6}{rgb}{0.75,0,0.25}
\definecolor{BTORMIX7}{rgb}{0.875,0,0.125}

\definecolor{BTOGMIX1}{rgb}{0, 0.125, 0.875}
\definecolor{BTOGMIX2}{rgb}{0, 0.25, 0.75}
\definecolor{BTOGMIX3}{rgb}{0, 0.375, 0.625}
\definecolor{BTOGMIX4}{rgb}{0, 0.50, 0.50}
\definecolor{BTOGMIX5}{rgb}{0, 0.625, 0.375}
\definecolor{BTOGMIX6}{rgb}{0, 0.75 ,0.25}
\definecolor{BTOGMIX7}{rgb}{0, 0.875, 0.125}

\begin{document}
%
\title{Achievable Rate and Energy Efficiency of Hybrid and Digital Beamforming Receivers with Low Resolution ADC}
%
%
%

\author{Kilian~Roth,~\IEEEmembership{Member,~IEEE,}
        Josef~A.~Nossek,~\IEEEmembership{Life Fellow,~IEEE}%
\thanks{K. Roth is with Next Generation and Standards, Intel Deutschland GmbH, Neubiberg 85579, Germany (email: $\{$kilian.roth$\}$@intel.com)}%
\thanks{K. Roth and J. A. Nossek are with the Department of Electrical and Computer Engineering, Technical University Munich, Munich 80290, Germany (email: $\{$kilian.roth, josef.a.nossek$\}$@tum.de)}%
\thanks{J. A. Nossek is with Department of Teleinformatics Engineering, Federal University of Ceara, Fortaleza, Brazil}}%

\maketitle

\input{./Introduction/Introduction}

\input{./SignalModel/SignalModel}

\input{./PowerModel/PowerModel}

\input{./AchievableRateExpressions/AchievableRateExpressions}

\input{./AchievableRateComparison/AchievableRateComparison}
\input{./conclusion/conclusion}

\input{./appendix/appendix}

\input{./literature/literature}

\input{./biography/biography}

\end{document}

%% file: Introduction/Introduction.tex
\begin{abstract}
For 5G it will be important to leverage the available millimeter wave spectrum.
To achieve an approximately omnidirectional coverage with a similar effective antenna aperture compared to state of the art cellular systems, an antenna array is required at both the mobile and basestation. 
Due to the large bandwidth, the analog front-end of the receiver with a large number of antennas becomes especially power hungry.
Two main solutions exist to reduce the power consumption: Hybrid BeamForming (HBF) and Digital BeamForming (DBF) with low resolution Analog to Digital Converters (ADCs).
Hybrid beamforming can also be combined with low resolution ADCs. 
This paper compares the spectral and energy efficiency based on the RF-frontend configuration. 
A channel with multipath propagation is used. 
In contrast to previous publication, we take the spatial correlation of the quantization noise into account. 
We show that the low resolution ADC digital beamforming is robust to small Automatic Gain Control (AGC) imperfections. 
We showed that in the low SNR regime the performance of DBF even with 1-2 bit resolution outperforms HBF. If we consider the relationship of spectral and energy efficiency, DBF with 3-5 bits resolution achieves the best ratio of spectral efficiency per power consumption of the RF receiver frontend over a wide SNR range. The power consumption model is based on components reported in literature.
\end{abstract}
%
\IEEEpeerreviewmaketitle
%
%
%
%
\begin{figure*}[!t]
\begin{center}
\normalsize
	\psfrag{MT}[][]{$M_T$}
	\psfrag{MA}[][]{$M_{C}$}
	\psfrag{MR}[][]{$M_{RFE}$}
	\ifCLASSOPTIONdraftcls
		\psfrag{transmitter}[0][0][1][180]{transmitter}
	\else
		\psfrag{transmitter}[0][0]{transmitter}
	\fi
	\psfrag{MIMO}[][]{MIMO}
	\psfrag{channel}[][]{channel}
	\psfrag{analog}[][]{analog}
	\psfrag{signal}[][]{signal}
	\psfrag{combiner}[][]{combination}
	\psfrag{A/D}[][]{A/D}
	\psfrag{conversion}[][]{conversion}
	\psfrag{digital}[][]{digital}
	\psfrag{baseband}[][]{baseband}
	\centering
	\ifCLASSOPTIONdraftcls
		\includegraphics{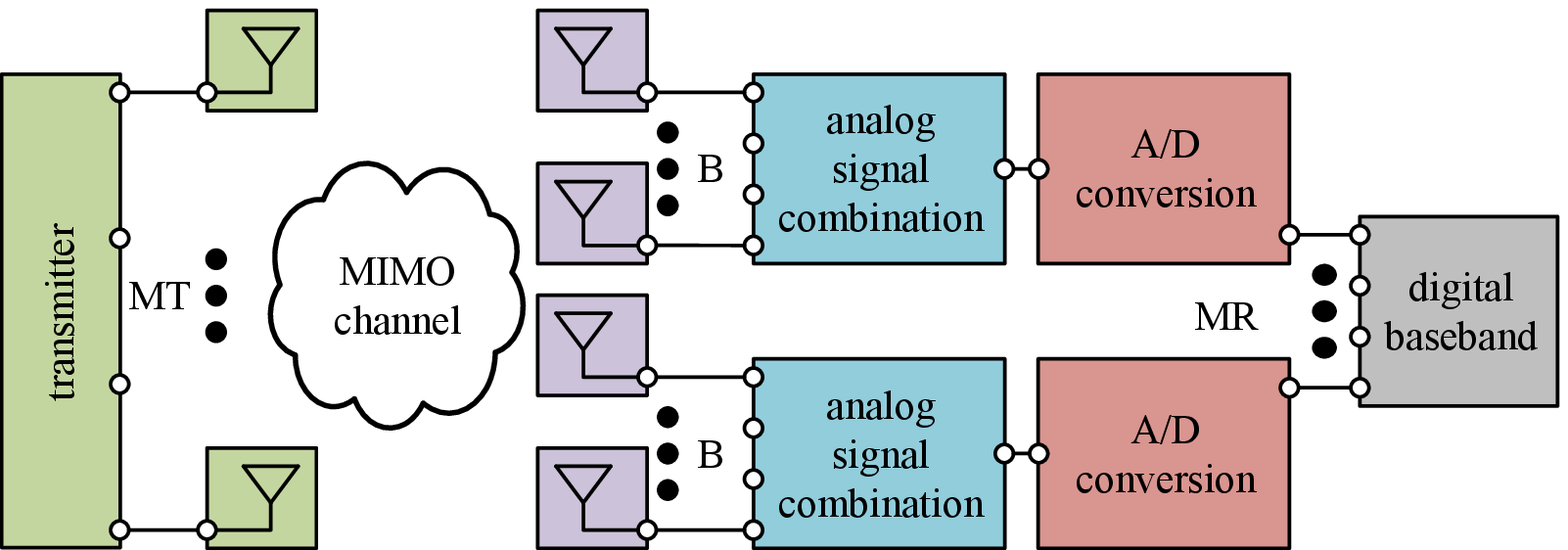}
	\else
		\includegraphics{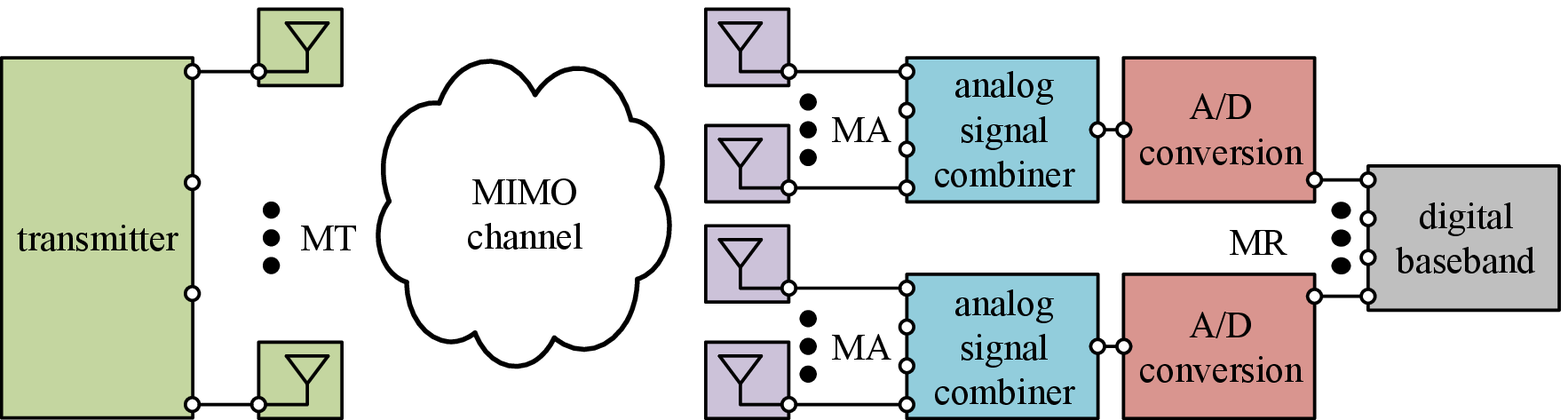}
	\fi
	\caption{System model with $M_T$ transmit antennas and $M_{C}$ antennas at each of the $M_{RFE}$ RF chains. Number of receive antennas $M_R$ is equal to $M_{C} \cdot M_{RFE}$.}
	\label{fig:SystemModel}
\hrulefill
\end{center}
\vspace*{4pt}
\end{figure*}
\section{Introduction}
The use of the available bandwidth in the frequency range of 6 to 100 GHz is considered to be an essential part of the next generation mobile broadband standard 5G \cite{FIVEDISRUPTIVE}. 
Due to the propagation condition, this technology is especially attractive for high data rate, low range wireless communication. 
This frequency range is referred to as millimeter wave (mmWave), even though it contains the lower 
centimeter wave range. In the last years, the available spectrum and the start of the availability of consumer grade systems lead to a huge increase in academic and industrial research. 
However, to fully leverage the spectrum while being power-efficient, the Base Band (BB) and Radio Front-End (RFE) capabilities must be drastically changed from state of the art cellular devices. 

The use of high carrier frequencies above 6 GHz will go hand in hand with the implementation of large antenna arrays 
\cite{FIVEDISRUPTIVE}, \cite{WHATWILL5GBE}. 
The support of a large number of antennas at the mobile and base station requires a new frontend design.
To attain a similar link budget, the effective antenna aperture of a mmWave system must be comparable to current
systems operating at carrier frequencies below 6 GHz. Therefore, an antenna array at 
the base and mobile station is unavoidable. Since the antenna gain and thus the directivity 
increases with the aperture, an antenna array is the only solution to achieve a high effective aperture while maintaining an omnidirectional coverage. 

\subsection{Related Work}
Current LTE systems have a limited amount of antennas at the base and mobile stations. Since the bandwidth is relatively narrow,
the power consumption of having a receiver RF chain with a high resolution ADC at each antenna is still feasible.
For future mmWave mobile broadband systems, a much larger bandwidth \cite{NGNM5GWHITE} and a large number of antennas
are being considered \cite{FIVEDISRUPTIVE}. 
The survey \cite{ADCSURVEY} shows that ADCs
with an extensive sampling frequency, and medium number of effective bits consume a considerable amount of power.
The ADC can be considered as the bottleneck of the receiver \cite{1BITMIMOINFO1}. 

The antenna array combined with the large bandwidth is a huge challenge for the hardware implementation, essentially
the power consumption will limit the design space. At the moment, analog or hybrid beamforming are considered 
as a possible solution to reduce the power consumption.
Analog or hybrid beamforming systems highly depend on the calibration of the analog components. Another major disadvantage is the 
large overhead associated with the alignment of the Tx and Rx beams of the base and mobile station. 
Specifically, if high gain is needed, the beamwidth is small and thus the acquisition and constant alignment of the optimal beams in a 
dynamic environment is very challenging \cite{CellSearchDirectionalmmW, rappaport2014millimeter, 1BITCAPHEATH}.

The idea of hybrid beamforming is based on the concept of phase array antennas commonly used in radar application \cite{MAILLOUXPHASEDARRAY}. 
Due to the reduced power consumption, it is also seen as a possible solution for mmWave mobile broadband communication\cite{HBFMAG}. 
If the phase array approach is combined with digital beamforming the phase array approach might also be feasible for non-static or quasi static scenarios.
In \cite{Kong:EECS-2014-191}, it was shown that considering the inefficiency of mmWave amplifiers and the high insertion loss of RF phase shifters, it
is better to perform the phase shifting in the baseband. The power consumption of both cases is comparable, as long as the 
number of antennas per RF-chain remains relatively small. 

Another option to reduce the power consumption while keeping the number of antennas constant is, to reduce the power consumption of the ADCs by
reducing their resolution. This can also be combined with hybrid beamforming. Some of these evaluations consider only the extreme case of 1-bit quantization 
\cite{1BITCAPHEATH, 1BITIDEA2, 1BITMIMOINFO1}. In \cite{AQNMAMINE, QINGOPAQNM} the Analog to Digital (A/D) conversion is modeled as a linear
stochastic process. Low resolution A/D conversion combined with OFDM in an uplink scenario are considered in \cite{ADCOFDM, OFDM1BIT}.

In \cite{HLOWADC, ACHCDCCOMP} hybrid beamforming with low resolution A/D conversion was considered.
The energy efficiency / spectral efficiency trade-off of fully-connected hybrid and digital beamforming with low resolution ADCs is assessed in 
\cite{ACHCDCCOMP}. But in contrast as shown in the system diagram in Figure \ref{fig:SystemModel}, we consider
a hybrid beamforming system that has exclusive antennas per RF-chain (aka. sub-array hybrid beamforming).
In this work we concentrated on effects of the hardware constraints at the receiver, thus we assumed the transmitter to be ideal.
In \cite{ACHCDCCOMP}, a fully-connected hybrid beamforming system is used, this has a large
additional overhead
 associated with an increased number of phase shifter and larger power combiners. Also in this case additional amplifiers to compensate for the insertion-loss of the RF phase shifters and combiners are required.
In \cite{ONERLOWRES}, analog beamforming is compared with digital beamforming in terms of power efficiency.

\subsection{Contribution}
In this paper, we assess the achievable rate of hybrid and digital beamforming with low resolution A/D conversion in a multipath environment. 
The paper \cite{ACHCDCCOMP} showed that a digital beamforming system is always more energy efficient than a fully-connected hybrid beamforming system. 
In contrast we use a hybrid beamforming system with exclusive antennas, which has a greatly reduced hardware complexity compared to fully-connected hybrid beamforming.
Therefore, in our evaluation different systemas are compared. 
\begin{itemize}
	\item The achievable rate of hybrid and digital beamforming with low resolution ADC in a multipath environment is derived. The phase shifters of hybrid beamforming are not frequency selective, therefore if considering a comparison between hybrid and digital beamforming it is important to consider multipath channels.
The evaluation shows that the digital beamforming system for any resolution of the ADC always outperforms the hybrid system in the low SNR regime. It is important to stress that the low per antenna 
SNR regime is very likely the practical operating point of future mmWave systems. The low resolution ADC is essentially limiting the performance in the high SNR regime. Therefore, a hybrid system with higher resolution ADC will always at some point surpass the digital system with lower resolution. We also show that small imperfections in the AGC do not degrade the performance of the digital system. 
	\item By including the off-diagonal elements of the quantization-error covariance matrix, the Additive Quantization Noise Model (AQNM) is refined in this work. For a scenario with very low resolution ADC (1-2 bit) and a larger number of receive antennas than transmit antennas, it is important to take this off-diagonal elements into account. 
	\item Energy efficiency and spectral efficiency of the given systems are characterized. We show that for a wide SNR range the digital beamforming system is more energy efficient than the hybrid beamforming one. We also show that an A/D resolution in the range of 3-5 lead to the most energy efficient receiver. 
\end{itemize}

\subsection{Notation}
Throughout the paper we use boldface lower and upper case letters to represent column vectors and matrices.
The term $a_{m,l}$ is the element on row $m$ and column $l$ of matrix $\boldsymbol{A}$ and $a_m$ is the $m$th element of vector $\boldsymbol{a}$. 
The expressions $\boldsymbol{A}^*$, $\boldsymbol{A}^T$, $\boldsymbol{A}^H$, and $\boldsymbol{A}^{-1}$ 
represent the complex conjugate, the transpose, the Hermitian, and the inverse of the matrix $\boldsymbol{A}$.
The symbol $\boldsymbol{R}_{\boldsymbol{a}\boldsymbol{b}}$ is the correlation matrix of vector $\boldsymbol{a}$ and $\boldsymbol{b}$ defined as
$\mathbb{E}[\boldsymbol{a}\boldsymbol{b}^H]$.
The Discrete Fourier Transformation (DFT) $\mathcal{F}(\cdot)$ and its inverse $\mathcal{F}^{-1}(\cdot)$ and the Fourier transformation $\mathscr{F}\{\cdot\}$ and its inverse
$\mathscr{F}^{-1}\{\cdot\}$ are also used.

%% file: SignalModel/SignalModel.tex
\section{System Model}
\subsection{Signal Model}
\begin{figure}
	\centering
	\ifCLASSOPTIONdraftcls
		\input{SignalModel/pics/SignalModelpsfrag_column1}
	\else
		\input{SignalModel/pics/SignalModelpsfrag}
	\fi
	\caption{Signal Model.}
	\label{fig:SignalModel}
\end{figure}
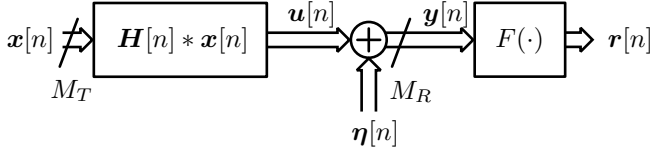

The signal model is shown in Figure \ref{fig:SignalModel}. The symbols $\boldsymbol{x}[n]$, 
$\boldsymbol{H}[n]$, $\boldsymbol{\eta}[n]$, and $\boldsymbol{y}[n]$ represent the transmit
signal, channel, noise, and receive signal of a system at time $n$. $M_T$ transmit and $M_R$ receive antennas are used. 
Since we assume a channel with multipath propagation the receive signal $\boldsymbol{y}[n]$ is defined as:
\begin{equation}
	\boldsymbol{y}[n] = \sum^{L-1}_{l=0} \boldsymbol{H}[l] \boldsymbol{x}[n-l] + \boldsymbol{\eta}[n],
\end{equation}
where $L$ is the maximum delay of the channel in samples.
The operation $F(\cdot)$ is defined as multiplication with the analog receiver beamforming matrix $\boldsymbol{W}_R$
followed by a quantization operation $Q_b(\cdot)$ with resolution of $b$ bits:
\begin{equation}
	\boldsymbol{r}[n] = F(\boldsymbol{y}[n]) = Q_b(\boldsymbol{y}_C[n]) = Q_b(\boldsymbol{W}_R^H \boldsymbol{y}[n]).
\end{equation}

We restricted the system to have $M_{C}$ antennas
exclusively connected to one RF front-end chain (see Figure \ref{fig:SystemModel}). Therefore, the matrix $\boldsymbol{W}_R$ has the form:
\begin{equation}
	\boldsymbol{W}_R = 
	\begin{bmatrix}
		\boldsymbol{w}_R^1  & \boldsymbol{0}_{M_C}   & \hdots  & \boldsymbol{0}_{M_C}  \\
		\boldsymbol{0}_{M_C} & \boldsymbol{w}_R^2  & \ddots  & \boldsymbol{0}_{M_C}  \\
		\vdots & \ddots  & \ddots &\vdots    \\
		\boldsymbol{0}_{M_C} & \hdots  & \boldsymbol{0}_{M_C} & \boldsymbol{w}_R^{M_{RFE}} \\
	\end{bmatrix} \in \mathbb{C}^{M_R\times M_{RFE}},
\end{equation}
where the vector $\boldsymbol{w}_R^i$ is the analog beamforming vector of the $i$th RF chain. 
We also restrict our evaluation to each RF chain utilizing the same number of antennas $M_{C}$. 
The vectors $\boldsymbol{w}_R^i$ and $\boldsymbol{0}_{M_C}$ have dimension $M_{C}$. 

The use of analog beamforming is envisioned in many future mobile broadband systems, 
especially in the mmWave frequency range (\cite{MMWMIMOBFSM, ABFPS}). 
Since the complete channel matrix cannot be directly observed, one practical solution is scanning
different spatial direction (beams) and then select the configuration maximizing the SNR. There are
many different possibilities for selecting the optimal beam, e.g. 802.11ad is using a procedure based on exhaustive search \cite{WIGIGSTD}.

For the evaluation, we assume that the antennas of each RF chain form a Uniform Linear Array (ULA). 
If a planar wavefront is impinging on the ULA and the spacing of adjacent antennas is $d = \lambda / 2$, the receive signal at adjacent antennas 
is phase shifted by $\phi_i = \pi  \sin(\theta_i)$.
The angle $\theta_i$ is the angle of a planar wavefront relative to the antennas of the ULA. 
This formula assumes that a planar wavefront is impinging at the antenna array, and that the symbol duration is large relative to the maximum delay between two antennas. 
With the constraint of observing only a single spatial direction, the receive vector $\boldsymbol{w}^i_R$ for an ULA antenna array takes the form:
\begin{equation}
	\boldsymbol{w}^i_R = \left[1,e^{j \phi_i},e^{j2 \phi_i},\cdots,e^{j(M_{C} - 1) \phi_i}\right]^H.
\end{equation}
In the special case of full digital beamforming ($M_{C} = 1$ and therefore $M_{RFE} = M_R$), $\boldsymbol{W}_R$ is
equal to the identity matrix $\boldsymbol{I}$ of size $M_R \times M_R$. 

The quantization operator $Q_b(\boldsymbol{a})$ is treating the I and Q component of each element of a vector $\boldsymbol{a}$ separately.
For a real valued, scalar input $a$, the output of the operation is defined as:
\begin{equation}
	r = Q_b(a) = q^j ~\forall ~a ~\in \left]q_l^{j-1} q_l^{j}\right].
\end{equation}
Here $q^j$ is the representative of the $j$th quantization bin with the input interval $\left]q_l^{j-1} q_l^{j}\right]$. To cover a real valued input
the left limit of the first interval $q_l^{0}$ and the right limit of the last interval $q_l^{N_b}$ are equal to $-\infty$ and $\infty$ respectively. 
The number of quantization bins $N_b$ is equal to $2^b$.
For real world ADC the difference between representatives of quantization bins $q_j$ and the size of the quantization bins are uniform.
We thus limit our evaluation to this set of quantizers. For the theoretical evaluation we assume Gaussian signaling. Consequently, we use the stepsize to minimize the distortion for Gaussian
signals shown in \cite{QUANTMAX}.

Since the actual receive power at each antenna can be different, an AGC needs to adapt a Variable Gain Amplifier (VGA) to generate the 
minimal distortion. To simplify our model, we assume that the AGC is always perfectly adapting to the current situation. Since in practice an AGC cannot accomplish this task without
error, we will show the impact of an imperfect AGC. We model this by a relative error to the perfect gain value.

For the rest of the paper we define the SNR $\gamma$ as:
\begin{equation}
	\gamma = \frac{\mathbb{E} \left[ \left\vert \left\vert \sum\limits^{L-1}_{l=0} \boldsymbol{H}[l] \boldsymbol{x}[n-l] \right\vert \right\vert^2_2 \right]}{ \mathbb{E} [\left\vert \left\vert \boldsymbol{\eta}[n] \right\vert \right\vert^2_2 ]}.
\end{equation}
This formula is basically just describing the average SNR at each antenna. It is important to note that the expectation takes the realization of the channel and realizations of $\boldsymbol{x}[n]$ into account.

\subsection{Channel Model}
Dependent on the scenario, different channel models are used:
\begin{itemize}
	\item Finite path model with all paths arriving at the same time
	\item Finite path model with exponential Power Delay Profile (PDP)
\end{itemize}
The channel models assume different rays impinging on the receiver antenna array. In the first example, they are assumed to arrive at the receiver antennas at the same time.
Under the assumption of a ULA at the transmitter and receiver, a channel consisting of $K$ different rays can be modeled as:
\begin{equation}
	\boldsymbol{H} = \frac{1}{\sqrt{K M_T}}\sum_{k = 1}^{K} \alpha(k) \boldsymbol{a}_r(\phi_r(k)) \boldsymbol{a}_t^T(\phi_t(k)).
\end{equation}
The vectors $\boldsymbol{a}_r(\phi_r(k))$ and $\boldsymbol{a}_t(\phi_t(k))$ are the array steering vectors at the receiver and transmitter. 
The phase shift between
the signal of adjacent antenna elements $\phi_r(k)$ and $\phi_t(k)$ of path $k$ depend on the angle of arrival $\theta_r(k)$ and departure $\theta_t(k)$ .
\begin{equation}
	\boldsymbol{a}_r^T(\phi_r(k)) = \left[1,e^{j \phi_r(k)},e^{j2 \phi_r(k)},\cdots,e^{j(M_r - 1) \phi_r(k)}\right].
\end{equation}
The transmit vectors $\boldsymbol{a}_t^T(\phi_r(k))$ has the same form as $\boldsymbol{a}_r^T(\phi_r(k))$. 
The complex gains $\alpha(k)$ are circular symmetric Gaussian distributed with zero mean and unit variance.
Except for the different normalization factory, this channel model is the same as the one presented in \cite{Mo2014}.
The difference comes from the fact that the sum power of the transmit signal is constraint to be less or equal to $M_T$.
To set the average per antenna receive power to one we normalize the channel by $\frac{1}{\sqrt{K M_T}}$.
The angles of arrival $\theta_r(k)$ and departure $\theta_t(k)$ are uniformly distributed in the range of  $-\pi$ to $\pi$. 

Since in real world scenario the different rays are reflection of different scatterers, the path of each of these rays from the transmitter to the receiver has a different
length. This results in rays arriving at the receiver at different time. In a simplified case, it can be expected that the path that arrives at a later time have a lower power.
The measurements in \cite{mmMAGIC_D2_1} show that for channels at 60 GHz an exponential Power Delay Profile (PDP) is sufficiently approximating a real world scenario.
\begin{equation}
	\boldsymbol{H}[l] = \frac{1}{\sqrt{M_T} } \alpha(l) \boldsymbol{a}_r(\phi_r(l)) \boldsymbol{a}_t^T(\phi_t(l)).
\end{equation}
Here we assume, that at delay $l$ only one ray arrives at the receiver. Here the complex gain of the ray $\alpha(l)$ is circular symmetric Gaussian distributed with zero mean and a variance defined according to:
\begin{equation}
	v_l = \mathbb{E}\left[\left\vert\alpha(l)\right\vert^2\right]  = e^{-\beta l}.
\end{equation}
The parameter $\beta$ defines the how fast the power decays in relation to the sample time.
The additional parameters are the maximum channel length in samples $L$ and the number of present channel tabs $P$. 
This means that for all possible present channel rays $\boldsymbol{v}$ of length $L$, $P$ positions are selected for each channel realization. At all other positions, $\boldsymbol{v}$ is equal to $0$.
To normalize the average power, the variance vector $\boldsymbol{v}$ is normalized by:
\begin{equation}
	\boldsymbol{v}_n = \frac{\boldsymbol{v}}{\vert\vert\boldsymbol{v}\vert\vert^2}.
\end{equation}

%% file: SignalModel/pics/SignalModelpsfrag_column1.tex
\begingroup%
  \makeatletter%
  \providecommand\color[2][]{%
    \errmessage{(Inkscape) Color is used for the text in Inkscape, but the package 'color.sty' is not loaded}%
    \renewcommand\color[2][]{}%
  }%
  \providecommand\transparent[1]{%
    \errmessage{(Inkscape) Transparency is used (non-zero) for the text in Inkscape, but the package 'transparent.sty' is not loaded}%
    \renewcommand\transparent[1]{}%
  }%
  \providecommand\rotatebox[2]{#2}%
  \ifx\svgwidth\undefined%
    \setlength{\unitlength}{308.62999228bp}%
    \ifx\svgscale\undefined%
      \relax%
    \else%
      \setlength{\unitlength}{\unitlength * \real{\svgscale}}%
    \fi%
  \else%
    \setlength{\unitlength}{\svgwidth}%
  \fi%
  \global\let\svgwidth\undefined%
  \global\let\svgscale\undefined%
  \makeatother%
  \begin{picture}(1,0.23526553)%
    \put(0,0){\includegraphics[width=\unitlength]{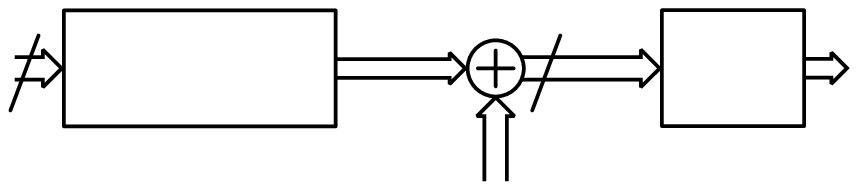}}%
    \put(0.75441046,0.15552104){\color[rgb]{0,0,0}\makebox(0,0)[lb]{\smash{$F(\cdot)$}}}%
    \put(0.03483135,0.15552104){\color[rgb]{0,0,0}\makebox(0,0)[lb]{\smash{$\boldsymbol{x}[n]$}}}%
    \put(0.91468749,0.15552104){\color[rgb]{0,0,0}\makebox(0,0)[lb]{\smash{$\boldsymbol{r}[n]$}}}%
    \put(0.19696303,0.15552104){\color[rgb]{0,0,0}\makebox(0,0)[lb]{\smash{$\boldsymbol{H}[n]*\boldsymbol{x}[n]$}}}%
    \put(0.10036289,0.0824784){\color[rgb]{0,0,0}\makebox(0,0)[lb]{\smash{$M_T$}}}%
    \put(0.53931893,0.02002498){\color[rgb]{0,0,0}\makebox(0,0)[lb]{\smash{$\boldsymbol{\eta}[n]$}}}%
    \put(0.64374818,0.19426274){\color[rgb]{0,0,0}\makebox(0,0)[lb]{\smash{$\boldsymbol{y}[n]$}}}%
    \put(0.59634514,0.08069633){\color[rgb]{0,0,0}\makebox(0,0)[lb]{\smash{$M_R$}}}%
    \put(0.4460033,0.19426274){\color[rgb]{0,0,0}\makebox(0,0)[lb]{\smash{$\boldsymbol{u}[n]$}}}%
  \end{picture}%
\endgroup%

%% file: SignalModel/pics/SignalModelpsfrag.tex
\begingroup%
  \makeatletter%
  \providecommand\color[2][]{%
    \errmessage{(Inkscape) Color is used for the text in Inkscape, but the package 'color.sty' is not loaded}%
    \renewcommand\color[2][]{}%
  }%
  \providecommand\transparent[1]{%
    \errmessage{(Inkscape) Transparency is used (non-zero) for the text in Inkscape, but the package 'transparent.sty' is not loaded}%
    \renewcommand\transparent[1]{}%
  }%
  \providecommand\rotatebox[2]{#2}%
  \ifx\svgwidth\undefined%
    \setlength{\unitlength}{254.82999363bp}%
    \ifx\svgscale\undefined%
      \relax%
    \else%
      \setlength{\unitlength}{\unitlength * \real{\svgscale}}%
    \fi%
  \else%
    \setlength{\unitlength}{\svgwidth}%
  \fi%
  \global\let\svgwidth\undefined%
  \global\let\svgscale\undefined%
  \makeatother%
  \begin{picture}(1,0.23763686)%
    \put(0,0){\includegraphics[width=\unitlength]{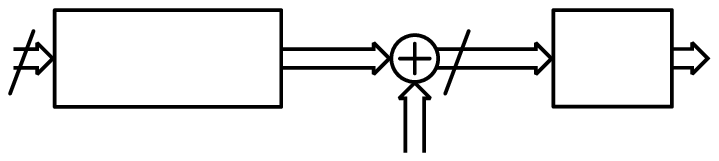}}%
    \put(0.75054891,0.15867983){\color[rgb]{0,0,0}\makebox(0,0)[lb]{\smash{$F(\cdot)$}}}%
    \put(0.0276655,0.15867983){\color[rgb]{0,0,0}\makebox(0,0)[lb]{\smash{$\boldsymbol{x}[n]$}}}%
    \put(0.91661107,0.15867983){\color[rgb]{0,0,0}\makebox(0,0)[lb]{\smash{$\boldsymbol{r}[n]$}}}%
    \put(0.19082349,0.15867983){\color[rgb]{0,0,0}\makebox(0,0)[lb]{\smash{$\boldsymbol{H}[n]*\boldsymbol{x}[n]$}}}%
    \put(0.09280697,0.08390849){\color[rgb]{0,0,0}\makebox(0,0)[lb]{\smash{$M_T$}}}%
    \put(0.53761331,0.02063101){\color[rgb]{0,0,0}\makebox(0,0)[lb]{\smash{$\boldsymbol{\eta}[n]$}}}%
    \put(0.6430954,0.19712122){\color[rgb]{0,0,0}\makebox(0,0)[lb]{\smash{$\boldsymbol{y}[n]$}}}%
    \put(0.59384688,0.0821426){\color[rgb]{0,0,0}\makebox(0,0)[lb]{\smash{$M_R$}}}%
    \put(0.44323667,0.19712122){\color[rgb]{0,0,0}\makebox(0,0)[lb]{\smash{$\boldsymbol{u}[n]$}}}%
  \end{picture}%
\endgroup%

%% file: PowerModel/PowerModel.tex
\subsection{Power Consumption Model}
\label{sec:power}
In a future 5G millimeter Wave mobile broadband system, it will be necessary to utilize large antenna arrays.
It is therefore important to compare the power consumption of different receiver architectures. 
In this section we present a power model for analog/hybrid beamforming and digital beamforming in 
combination with low resolution ADCs. 

Since the spectrum in the 60 GHz band can be accessed without a license, it got significant attention.
Especially the WiGig (802.11ad) standard operating in this band, increased the transceiver RF hardware R\&D activities. 
Many chips were reported from industry and academia. Thus, 
it is safe to assume that the design reached a certain maturity, and performance figures derived from them represent 
the performance that is possible for a low cost CMOS implementation today. 

According to the discussion in \cite{Chen:EECS-2014-42}, baseband or IF phase shifting in contrast to RF phase shifting is assumed. 
This has the advantage of increased accuracy, decreased insertion loss, and reduced gain mismatch.
In \cite{Chen:EECS-2014-42}, the authors showed that the power consumption for a low number of antennas per RF-chain
is equivalent to a system utilizing RF Phase Shifters (PS). 

\begin{figure}
	\psfrag{LNA}[][]{LNA}
	\psfrag{mixers}[][]{mixers}
	\psfrag{0}[][]{$0^{\circ}$}
	\psfrag{90}[][]{$90^{\circ}$}
	\psfrag{LO}[][]{LO}
	\psfrag{hybrid and}[][]{hybrid and}
	\psfrag{LO buffer}[][]{LO buffer}
	\centering
	\includegraphics{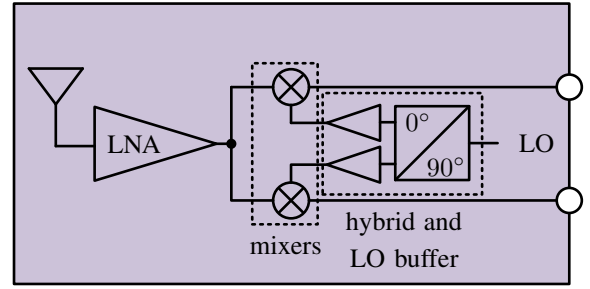}
	\caption{Direct conversion receiver.}
	\label{fig:powerfig100}
\end{figure}

\begin{figure}
	\psfrag{phase}[][]{phase}
	\psfrag{shifter}[][]{shifter}
	\psfrag{analog}[][]{analog}
	\psfrag{combiner}[][]{combiner}
	\psfrag{MA}[][]{$M_{C}$}
	\centering
	\includegraphics{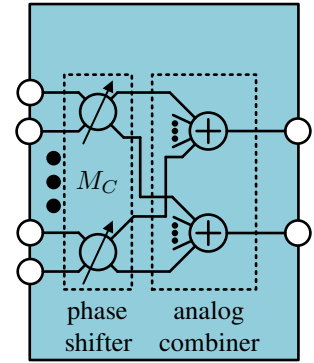}
	\caption{Analog signal combination.}
	\label{fig:powerfig200}
\end{figure}

All systems utilize the same direct conversion receiver (Figure \ref{fig:powerfig100}) to convert the signal into the analog baseband. For each system, we assume that the Local Oscillator (LO) is
shared by the whole system.
For the case of analog/hybrid beamforming systems, the analog baseband signals are phase shifted and then combined to generate the input signal of the $M_{RFE}$ ADCs (Figure \ref{fig:powerfig200}).

\begin{figure}
	\psfrag{VGA}[][]{VGA}
	\psfrag{ADC}[][]{ADC}
	\centering
	\includegraphics{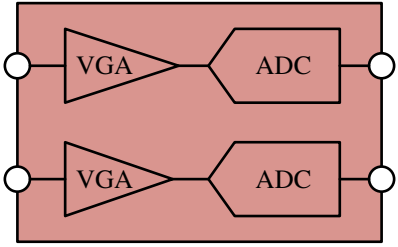}
	\caption{A/D conversion.}
	\label{fig:powerfig300}
\end{figure}

The A/D conversion consists of a VGA that is amplifying the signal to use the full dynamic range of the ADC (Figure \ref{fig:powerfig300}). 
For the special case of 1-bit quantized digital beamforming a
VGA is not necessary. It can be replaced by a much simpler Limiting Amplifier (LA). 

The power consumption of each component, including a reference, are shown in Table \ref{tab:powertab100}.
An LO with a power consumption as low as 22.5 mW is reported in \cite{QVCOREF}.  
The power consumption of a LNA, a mixer including a quadrature-hybrid coupler, and a VGA are reported in \cite{WIGIGRECEIVER} as 5.4, 0.5, and 2 mW. 
The 90$^{\circ}$ hybrid and the LO buffer reported in \cite{Marcu:EECS-2011-132} have a combined power consumption of 3 mW. 
The power consumption of the mixer reported in \cite{MIXERREF} is as low as 0.3 mW. The survey in \cite{ADCSURVEY} gives a
good overview of state of the art ADCs regarding Effective Number Of Bits (ENOB), sampling rate, and power consumption. 
Taking the predicted curve for the Walden figure of merit in \cite{ADCSURVEY} for a sampling frequency of 2.5 GS/s, we get 15 fJ per conversion step.
A LA that consumes 0.8 mW is reported in \cite{LAEXAMPLE1}. In the 1-bit quantized system, the LA (aka. Schmitt trigger) 
is already producing a digital signal, 
therefore the 1-bit ADC can be replaced by a
flip flop (FF). The power consumption of a FF is negligible compared to the rest of the RF front-end. 
\begin{table}
	\renewcommand{\arraystretch}{1.3}
	\caption{Components with power consumption.}
	\label{tab:powertab100}
	\centering
		\begin{tabularx}{0.95\columnwidth}{|X|X|X|X|}
			\hline
			label & component & power consumption & reference \\ \hline \hline
			$P_{LO}$ & LO & 22.5 mW & \cite{QVCOREF}\\ \hline
			$P_{LNA}$ & LNA & 5.4 mW & \cite{WIGIGRECEIVER} \\ \hline
			$P_{M}$ & mixer & 0.3 mW & \cite{MIXERREF} \\ \hline
			$P_{H}$ & 90$^{\circ}$ hybrid and LO buffer & 3 mW & \cite{Marcu:EECS-2011-132} \\ \hline
			$P_{LA}$ & LA & 0.8 mW& \cite{LAEXAMPLE1} \\ \hline
			$P_{1}$ & 1-bit ADC & 0 mW & \\ \hline
			$P_{PS}$ & phase shifter & 2 mW& \cite{Kong:EECS-2014-191, Chen:EECS-2014-42}\\ \hline
			$P_{VGA}$ & VGA & 2 mW & \cite{WIGIGRECEIVER} \\ \hline
			$P_{ADC}$ & ADC & $15~\mu \text{W/GHz}$ $\cdot f_s 2^{\text{ENOB}}$   & \cite{ADCSURVEY} \cite{Chan2015} \cite{Xu2016} \\ \hline
		\end{tabularx}
\end{table}

With the power consumption of the components, it is possible to compute the power consumption of the overall receiver front-end $P_R$ as:
\begin{equation}
	\begin{gathered}
		P_{R} =  P_{LO} + M_R\left(P_{LNA} + P_{H} + 2P_{M}\right) +\\
		\text{flag}_{C}\left(M_R P_{PS}\right)  +\\
		M_{RFE}\left(  \neg \text{flag}_{1\text{bit}}\left(2P_{VGA} + 2P_{ADC}\right) + \text{flag}_{1\text{bit}}\left( 2P_{LA} \right)\right),
	\end{gathered}
\end{equation}
where $\text{flag}_{C}$ is indicating if analog combining is used:
\begin{equation}
		\text{flag}_{C} = \left\{\begin{array}{ll} 0, & M_{RFE} = M_R, M_C = 1 \\
         1, & \text{else}\end{array}\right. .
\end{equation}
The variable $\text{flag}_{1\text{bit}}$ is indicating if 1 or multibit quantization is used. 
The operator $\neg$ represents a logic negation. 
In the case of 1-bit quantization, the power consumption of the VGA is replaced by the one of the LA and the power consumption of the
1-bit quantizer is neglected with the above stated reasoning. This formula now contains all special cases of digital beamforming ($M_{RFE} = M_R$),
analog beamforming ($M_R > 0 ~\text{and}~ M_{RFE} = 1$) and hybrid beamforming.

A receiver directly designed for the 1-bit quantization digital beamforming systems is very likely to reduce 
the power consumptions even further. Due to the 1-bit quantization at the end of the analog part of the receiver, the linearity required of the circuits
before is greatly reduced. This would enable specialized designs to improve the performance in terms of power consumption, which are not exploited in this work.

%% file: AchievableRateExpressions/AchievableRateExpressions.tex
\section{Achievable Rate Expressions}
\label{sect:achievablerateExpression}
In this subsection achievable rate expressions for different scenarios are derived. 
The different scenarios are any combination of  flat fading channel, multipath channel, hybrid beamforming and digital beamforming with low resolution A/D conversion.
In the case of hybrid beamforming, first the analog receive vectors are calculated. Afterwards, the system including the analog combining is treated as an equivalent digital
beamforming system. 
\subsection{Hybrid Beamforming Vectors}
To mimic the behavior of a spatial scan, we restricted the receive vectors $\boldsymbol{w}_R^i$ of the $i$th RF chain to Vandermonde vectors.
A practical system would have a set of predefined beamforming configuration that are scanned for every sub-array. To obtain the optimal results, all combination of beams need to be tested by the receiver. 
This is a combinatorial problem with size growing exponentially with the number of receiver RF-chains.
To make the problem feasible, the scan is performed separately for each receiver RF chain. 
This problem can be formulated as:
\begin{equation}
	\boldsymbol{w}^i_R(\hat{\phi}) = \arg\!\max_{\boldsymbol{w}^i_R(\phi_B)} \sum\limits_{l = 0}^{L-1}\left\vert\left\vert \boldsymbol{w}^i_R(\phi_B)^H \boldsymbol{H}^i[l]\right\vert\right\vert^2_2 \text{with}~ \phi_B \in \mathbb{B},
\end{equation}
with $\mathbb{B}$ being the set of all spatial direction $\phi_B$ that are scanned. 
The channel $\boldsymbol{H}^i[l]$ contains the $M_{C}$ rows of $\boldsymbol{H}[l]$ that belong to the antennas of the $i$th RF chain.

This procedure mimics the receive beam training in a practical system as described in \cite{80211ayBF}.
For this case the transmitter is sending a known reference sequence. The receiver tries different receiver beamforming configurations separately on each subarray $i$ and
records the achieved channel quality metric.
Afterwards, the configuration resulting in the best channel is selected. 
In this work such a procedure is emulated by selecting the receive beamforming vector resulting in the highest receive energy, based on the channel knowledge.
This procedure avoids lengthy numerical simulation of sequence detection with different configurations, but leads to the same beamformer configuration.

To select the values in $\mathbb{B}$, we first calculated the array factor of the antenna array. With this array factor, we then select the spacing of the values $\phi_B$ uniform from $0$ to $2\pi$. 
Here we assume isotropic minimum scattering antennas. For ULA with spacing $\lambda/2$, the absolute value of the normalized array factor is defined as \cite[page 294]{balanis2005antenna}:
\begin{equation}
	AF = \frac{1}{M_{C}}\left\vert \frac{\sin\left(M_{C}\frac{\pi}{2}\sin\left(\theta - \phi_B\right)\right)}{\sin\left(\frac{\pi}{2}\sin\left(\theta - \phi_B\right)\right)}\right\vert.
\end{equation}
That means that for actual arriving angle $\theta$ choosing $\phi_B = \theta$ is optimal. But this would mean that we have an infinite grid of $\phi_B$.
Assuming a single wavefront arriving at the receiver and an uniformly distributed angle of the arriving signal $\theta$, we get the following expression for the average error
 $\epsilon$:
 \begin{equation}
	\frac{2}{\Delta}\int\displaylimits^{\frac{\Delta}{2}}_0 1 - \frac{1}{M_{C}}\left\vert \frac{\sin\left(M_{C}\frac{\pi}{2}\sin\left(x\right)\right)}{\sin\left(\frac{\pi}{2}\sin\left(x\right)\right)}\right\vert dx \le \epsilon.
\end{equation}
Setting a maximum allowed $\epsilon$, we can solve the equation for the distance $\Delta$ between two angles in the set $\mathbb{B}$: 
 \begin{equation}
	\frac{2}{\Delta}\left(\int\displaylimits^{\frac{\Delta}{2}}_0 1 - \frac{1}{M_{C}}\left\vert \frac{\sin\left(M_{C}\frac{\pi}{2}\sin\left(x\right)\right)}{\sin\left(\frac{\pi}{2}\sin\left(x\right)\right)}\right\vert dx\right) - \epsilon  = 0.
	\label{eq::IntBisect}
\end{equation}
Equation \eqref{eq::IntBisect} can be solved by a bisection based procedure. In this case we select a lower bound $\Delta_l$ and an upper bound $\Delta_u$ for $\Delta$, these values are chosen in a way to ensure that the value that solves the equation is in between them. Afterwards, Equation \eqref{eq::IntBisect} is solved for $\Delta_l$, $\Delta_u$ and $(\Delta_l + \Delta_u) / 2$ by numeric integration. Based on the results of this function evaluation we select the bounds for the next iteration of the bisection method.
A table for some configurations and the minimum number of elements in $\mathbb{B}$ are shown Table \ref{tab:rate100}. 
It can be observed that for the given parameters, the minimum number of elements can be well approximated by $4M_{C}$. Therefore we select $4M_{C}$ elements uniform in the range from $0$ to $2\pi$ to represent the set $\mathbb{B}$. 
\begin{table}
	\renewcommand{\arraystretch}{1.3}
	\caption{Minimum number of beams necessary to achieve mean error $\epsilon = 0.1$.}
	\label{tab:rate100}
	\centering
		\begin{tabularx}{0.95\columnwidth}{|X|X|X|}
			\hline
			$M_{C}$ & minimum number of beams & approximation $4M_{C}$ \\ \hline \hline
			2 &  7 & 8 \\ \hline
			4 & 14 & 16 \\ \hline
			8 &  29 & 32 \\ \hline
			16 &  58 & 64 \\ \hline
			32 &  115 & 128 \\ \hline
		\end{tabularx}
\end{table}

After selecting all beamforming vectors $\boldsymbol{w}^i_R$, the overall matrix $\boldsymbol{W}_R$ is constructed.
With $\boldsymbol{W}_R$, we can generate the effective channel $\boldsymbol{H}_{C}[l]$:
\begin{equation}
	\boldsymbol{H}_{C}[l] = \boldsymbol{W}^H_R\boldsymbol{H}[l].
\end{equation}
and the effective noise covariance matrix $\boldsymbol{R}_{\boldsymbol{\eta}_{C}\boldsymbol{\eta}_{C}}$:
\begin{equation}
	\boldsymbol{R}_{\boldsymbol{\eta}_{C}\boldsymbol{\eta}_{C}} = \boldsymbol{W}^H_R \boldsymbol{R}_{\boldsymbol{\eta}\boldsymbol{\eta}} \boldsymbol{W}_R.
\end{equation}
The effective channel and noise covariance matrix are then input to the digital system with low resolution A/D conversion.
Algorithm \ref{alg:beamforming} shows the procedure of finding the receiver beamforming vectors $\boldsymbol{w}^i_R$ given the channel and the number of antennas per RF chain.
\begin{algorithm}
	\caption{Selection of the beamforming vectors.}
	\begin{algorithmic}[1]
		\Require{$\boldsymbol{H}[l]$, $M_{RFE}$ and $M_{C}$}
		\State $\mathbb{B} \gets \left\{\phi_1, \phi_2, \cdots, \phi_{4M_{C}}\right\}$
		\For{$i \gets 1 \textrm{ to } M_{RFE}$}
			\For{$j \gets 1 \textrm{ to } 4M_{C}$}
				\State $\boldsymbol{w}_{tst} \cdots \left[1, e^{\phi_j}, \cdots, e^{(M_{C}-1)\phi_j} \right]^H$
				\State $p^i(j) \gets \sum\limits_{l = 0}^{L-1}\left\vert\left\vert \boldsymbol{w}_{tst}^H \boldsymbol{H}^i[l]\right\vert\right\vert^2_2 $
			\EndFor
			\State $ \hat{j} \gets \arg\!\max\limits_j p^i(j)$
			\State $ \boldsymbol{w}^i_R \gets \left[1, e^{\phi_{\hat{j}}}, \cdots, e^{(M_{C}-1)\phi_{\hat{j}}} \right]^H$
		\EndFor
		\State \Return $\boldsymbol{w}^i_R~\forall i = \{1, \hdots, M_{RFE}\}$
  \end{algorithmic}
  \label{alg:beamforming}
\end{algorithm}

\subsection{Modeling the Quantization}
As the model used in \cite{AQNMAMINE, ONERLOWRES, HLOWADC}, we use the Bussgang theorem to decompose the signal after quantization in a signal component and an uncorrelated quantization error $\boldsymbol{e}$:
\begin{equation}
	\boldsymbol{r}[n] = \boldsymbol{F}\boldsymbol{y}_{C}[n] + \boldsymbol{e}[n],
\end{equation}
with $\boldsymbol{y}_{C}[n]$ being the signal after the analog combining at the receiver 
equal to $\boldsymbol{u}[n] + \boldsymbol{\eta}[n]$, where $\boldsymbol{u}[n]$ is the receive signal after the multipath channel. 
This is basically modeling the deterministic process of quantization as a random process.
The quantization distortion factor $\rho\left(Q_b(\cdot)\right)$ is defined as:
\begin{equation}
	\rho\left(Q_b(\cdot)\right) = \mathbb{E}\left[\left\vert a - Q_b(a)\right\vert^2\right].
\end{equation}
The input of the quantizer is assumed to be zero mean unit variance Gaussian random variable $a$. 
The distortion factor $\rho\left(Q_b(\cdot)\right)$ depends on the actual quantization operator $Q_b(\cdot)$ and represents the variance of the introduced distortion.
As shown in \cite{AQNMAMINE}, the matrix $\boldsymbol{F}$ can be calculated as:
\begin{equation}
	\boldsymbol{F} = \boldsymbol{R}_{\boldsymbol{r}\boldsymbol{y}_{C}}\boldsymbol{R}^{-1}_{\boldsymbol{y}_{C}\boldsymbol{y}_{C}}.
	\label{eq:matrixF1}
\end{equation}
With the definition of the distortion factor, this expression can be reduced to:
\begin{equation}
	\begin{gathered}
	\boldsymbol{R}_{\boldsymbol{r}\boldsymbol{y}_{C}} = \boldsymbol{R}_{\boldsymbol{y}_{C}\boldsymbol{r}} = \\
	 \left(1 - \rho\left(Q_b(\cdot)\right)\right) \text{diag}\left( \boldsymbol{R}_{\boldsymbol{y}_{C}\boldsymbol{y}_{C}}\right)^{-\frac{1}{2}} \boldsymbol{R}_{\boldsymbol{y}_{C}\boldsymbol{y}_{C}}.
	\end{gathered}
	\label{eq:matrixF2}
\end{equation}
Plugging \eqref{eq:matrixF2} into \eqref{eq:matrixF1} results in: 
 \begin{equation}
	\boldsymbol{F} = \left(1 - \rho\left(Q_b(\cdot)\right)\right) \text{diag}\left( \boldsymbol{R}_{\boldsymbol{y}_{C}\boldsymbol{y}_{C}}\right)^{-\frac{1}{2}}.
\end{equation}
The covariance matrix $\boldsymbol{R}_{\boldsymbol{e}\boldsymbol{e}}$ can be calculated from:
\begin{equation}
	\boldsymbol{R}_{\boldsymbol{e}\boldsymbol{e}} =  \boldsymbol{R}_{\boldsymbol{r}\boldsymbol{r}} -
 	\boldsymbol{R}_{\boldsymbol{r}\boldsymbol{y}_{C}}\boldsymbol{R}^{-1}_{\boldsymbol{y}_{C}\boldsymbol{y}_{C}}\boldsymbol{R}_{\boldsymbol{y}_{C}\boldsymbol{r}}.
	\label{eq:matrixRee1}
\end{equation}
In \cite{AQNMAMINE, QINGOPAQNM, HLOWADC, ACHCDCCOMP, ONERLOWRES} only the diagonal elements of $\boldsymbol{R}_{\boldsymbol{r}\boldsymbol{r}}$ are used. As we showed in \cite{PaperDSP2015} based on the assumption of Gaussian signaling, it is possible to calculate the
complete matrix $\boldsymbol{R}_{\boldsymbol{r}\boldsymbol{r}}$, which we will show changes the overall result. 
We define this operation of calculating $\boldsymbol{R}_{\boldsymbol{r}\boldsymbol{r}}$ from $\boldsymbol{R}_{\boldsymbol{y}_{C}\boldsymbol{y}_{C}}$ as $T(\boldsymbol{R}_{\boldsymbol{y}_{C}\boldsymbol{y}_{C}}, Q_b(\cdot))$ dependent on the quantization function $Q_b(\cdot)$. Using this definition and plugging \eqref{eq:matrixF2} into \eqref{eq:matrixRee1} we get:
\begin{equation}
	\begin{gathered}
	\boldsymbol{R}_{\boldsymbol{e}\boldsymbol{e}} =  T\left(\boldsymbol{R}_{\boldsymbol{y}_{C}\boldsymbol{y}_{C}}, Q_b(\cdot)\right) - \left(1 - \rho\left(Q_b(\cdot)\right)\right)^2 \\
	 \text{diag}\left( \boldsymbol{R}_{\boldsymbol{y}_{C}\boldsymbol{y}_{C}}\right)^{-\frac{1}{2}}\boldsymbol{R}_{\boldsymbol{y}_{C}\boldsymbol{y}_{C}} \text{diag}\left( \boldsymbol{R}_{\boldsymbol{y}_{C}\boldsymbol{y}_{C}}\right)^{-\frac{1}{2}}.
	\end{gathered}
	\label{eq:matrixRee2}
\end{equation}

Now we can calculated the effective channel $\boldsymbol{H}^\prime[l]$ and noise covariance matrix  $\boldsymbol{R}_{\boldsymbol{\eta}^\prime\boldsymbol{\eta}^\prime}$of the overall system including the analog combing and the quantization:
\begin{equation}
	\boldsymbol{H}^\prime[l] = \boldsymbol{F}\boldsymbol{W}_R^H\boldsymbol{H}[l],
\end{equation}
and
\begin{equation}
	\boldsymbol{R}_{\boldsymbol{\eta}^\prime\boldsymbol{\eta}^\prime} =  \boldsymbol{F}\boldsymbol{W}_R^H \boldsymbol{R}_{\boldsymbol{\eta}\boldsymbol{\eta}} \boldsymbol{W}_R \boldsymbol{F}^H +
	\boldsymbol{R}_{\boldsymbol{e}\boldsymbol{e}}.
	\label{eq:Rnndash}
\end{equation}
It is important to keep in mind that since $\boldsymbol{F}$ and $\boldsymbol{R}_{\boldsymbol{e}\boldsymbol{e}}$ are dependent on $\boldsymbol{R}_{\boldsymbol{y}\boldsymbol{y}}$, which dependents on
$\boldsymbol{R}_{\boldsymbol{x}\boldsymbol{x}}$, thus effective channel and noise covariance matrix changes with the transmit covariance matrix.
Up to now, the actual expression is exact for the case of a Gaussian input signal $\boldsymbol{x}$. The actual distribution of $\boldsymbol{e}$ is unknown. Approximating $\boldsymbol{e}$ by a Gaussian with the same covariance matrix leads 
to simple rate expressions and represents a worst case scenario.
For this statements we need to proof that $\mathbb{E}[\boldsymbol{u}[n]\boldsymbol{e}[n]]$ and $\mathbb{E}[\boldsymbol{\eta}[n]\boldsymbol{e}[n]]$ are equal to zero. 
With this choice of $\boldsymbol{F}$ we ensured that $\mathbb{E}[\boldsymbol{y}_C[n]\boldsymbol{e}^H[n]] = 0$.
We can expand this term to
\begin{equation}
	\mathbb{E}[\boldsymbol{y}_C[n]\boldsymbol{e}^H[n]] =  \mathbb{E}\left[\mathbb{E}[\boldsymbol{y}_C[n]\boldsymbol{e}^H[n] \vert \boldsymbol{\eta}[n]]\right] ,
\end{equation}
if we plug in the definition of $\boldsymbol{y}_C[n]$ we get  
\begin{equation}
	\mathbb{E}\left[\mathbb{E}[\boldsymbol{u}[n]\boldsymbol{e}^H[n]] + \boldsymbol{\eta}[n]\mathbb{E}[\boldsymbol{e}^H[n]]\right].
\end{equation}
Since the first term is independent of $\boldsymbol{\eta}[n]$ this reduces to
\begin{equation}
	\mathbb{E}[\boldsymbol{u}[n]\boldsymbol{e}^H[n]] + \mathbb{E}[\boldsymbol{\eta}[n]]\mathbb{E}[\boldsymbol{e}^H[n]].
\end{equation}
Since $\mathbb{E}[\boldsymbol{\eta}[n]]$ is equal to zero it follows that
\begin{equation}
	\mathbb{E}[\boldsymbol{u}[n]\boldsymbol{e}^H[n]] = 0.
\end{equation}
The proof for $\mathbb{E}[\boldsymbol{\eta}[n]\boldsymbol{e}^H[n]] = 0$ follows the same steps. 

\subsection{Calculation of the Receive Covariance Matrix}
For the calculation of the matrix $\boldsymbol{F}$ and the effective noise correlation matrix $\boldsymbol{R}_{\boldsymbol{\eta}^\prime\boldsymbol{\eta}^\prime}$, it
is necessary to calculate the correlation matrix $\boldsymbol{R}_{\boldsymbol{y}_{C}\boldsymbol{y}_{C}}$ of the signal after the analog combining. 
This signal is defined as:
\begin{equation}
	\begin{gathered}
		\boldsymbol{y}_{C}(t) = \boldsymbol{W}_R^H\left(\sum^{L-1}_{l=0} \boldsymbol{H}(l) \boldsymbol{x}(t-\tau_l) + \boldsymbol{\eta}(t)\right) \\
		= \boldsymbol{W}_R^H\left(\boldsymbol{u}(t) + \boldsymbol{\eta}(t) \right).
	\end{gathered}
\end{equation}
Since the two random variables $\boldsymbol{x}$ and $\boldsymbol{\eta}$ are independent the covariance matrix decomposes into:
\begin{equation}
	\boldsymbol{R}_{\boldsymbol{y}_{C}\boldsymbol{y}_{C}} = \boldsymbol{W}_R^H\left(\boldsymbol{R}_{\boldsymbol{u}(t)\boldsymbol{u}(t)} + \boldsymbol{R}_{\boldsymbol{\eta}(t)\boldsymbol{\eta}(t)} \right)\boldsymbol{W}_R.
\end{equation}
The remaining matrix that needs to be calculated is $\boldsymbol{R}_{\boldsymbol{u}(t)\boldsymbol{u}(t)}$.
\begin{equation}
	\boldsymbol{R}_{\boldsymbol{u}(t)\boldsymbol{u}(t)} = \mathbb{E}\left[\boldsymbol{u}(t)\boldsymbol{u}^H(t)\right].
\end{equation}
To simplify the notation for the derivation we evaluate the elements of the matrix separately:
\begin{equation}
	\left[\boldsymbol{R}_{\boldsymbol{u}(t)\boldsymbol{u}(t)}\right]_{i,j} = \mathbb{E}\left[u_i(t)u_j^*(t)\right].
\end{equation}
Without changing the result, we can transform this equation into the frequency domain and then transform it back. 
Since the expectation $\mathbb{E}$ and the Fourier transformation $\mathscr{F}$ are linear operations, we can interchange the order we perform them:
\begin{equation}
	\begin{gathered}
		\mathbb{E}\left[u_i(t)u_j^*(t)\right] = \mathscr{F}^{-1}\left\{\mathscr{F}\left\{\mathbb{E}\left[u_i(t)u_j^*(t)\right]\right\}\right\} \\
		= \mathscr{F}^{-1}\left\{\mathbb{E}\left[\mathscr{F}\left\{u_i(t)u_j^*(t)\right\}\right]\right\}.
	\end{gathered}
\end{equation}
With the convolution property of the Fourier transformation we get:
\begin{equation}
	\begin{gathered}
		\mathbb{E}\left[u_i(t)u_j^*(t)\right] = \mathscr{F}^{-1}\left\{\mathbb{E}\left[u_i(f)*u_j^*(-f)\right]\right\} \\
		= \mathscr{F}^{-1}\left\{\mathbb{E}\left[\int\limits_{-\infty}^{\infty}u_i(f^\prime)u_j^*(f^\prime-f)df^\prime\right]\right\}.
	\end{gathered}
\end{equation}
Due to the linearity of the expectation operation  $\mathbb{E}$ we can interchange it with the integrals:
\begin{equation}
		\mathbb{E}\left[u_i(t)u_j^*(t)\right] = \mathscr{F}^{-1}\left\{\int\limits_{-\infty}^{\infty}\mathbb{E}\left[u_i(f^\prime)u_j^*(f^\prime-f)\right]df^\prime\right\}.
\end{equation}
Since the transmit signals of each frequency bin $\boldsymbol{x}(f)$ are independent and have zero mean,
the expectation $\mathbb{E}\left[u_i(f^\prime)u_j^*(f^\prime-f)\right]$ is only unequal to zero if $f = 0$. 
Since outside of the transmission bandwidth the signal is going to be zero, we get the following expression:
\begin{equation}
		\mathbb{E}\left[u_i(t)u_j^*(t)\right] = \mathscr{F}^{-1}\left\{\delta(f)\int\limits_{f_1}^{f_2}\mathbb{E}\left[u_i(f^\prime)u_j^*(f^\prime)\right]df^\prime\right\},
\end{equation}
with the Dirac pulse $\delta(f)$ at frequency $f = 0$. So if we transform this Dirac impulse back into the time domain we get:
\begin{equation}
	\mathbb{E}\left[u_i(t)u_j^*(t)\right] = \int\limits_{f_1}^{f_2}\mathbb{E}\left[u_i(f)u_j^*(f)\right]df,
\end{equation}
independent of time $t$, that also states that we still have a stationary random process. 
Plugging the definition of $u_i(f)$ into the equation we get the covariance matrix $\boldsymbol{R}_{\boldsymbol{u}(t)\boldsymbol{u}(t)}$:
\begin{equation}
	\boldsymbol{R}_{\boldsymbol{u}(t)\boldsymbol{u}(t)} = \int\limits_{f_1}^{f_2}\boldsymbol{H}(f) \boldsymbol{R}_{\boldsymbol{x}(f)\boldsymbol{x}(f)}\boldsymbol{H}^H(f)df.
\end{equation}
Combining these results we can express the matrix $\boldsymbol{R}_{\boldsymbol{y}_{C}\boldsymbol{y}_{C}}$ as:
\begin{equation}
	\begin{gathered}
		\boldsymbol{R}_{\boldsymbol{y}_{C}\boldsymbol{y}_{C}} = \\
		\boldsymbol{W}_R^H\left(\int\limits_{f_1}^{f_2}\boldsymbol{H}(f) \boldsymbol{R}_{\boldsymbol{x}(f)\boldsymbol{x}(f)}\boldsymbol{H}^H(f)df + \boldsymbol{R}_{\boldsymbol{\eta}(t)\boldsymbol{\eta}(t)} \right)\boldsymbol{W}_R.
	\end{gathered}
	\label{eq:Ryycalc}
\end{equation}

\subsection{Problem Formulation}
For the given signal model, the problem of finding the maximum achievable rate for a multipath channel with full Channel State Information (CSI) at the Transmitter (CSIT) and the Receiver (CSIR) can be formulated as:
\begin{equation}
	\begin{gathered}
		R = \frac{1}{N}\max_{p(\boldsymbol{x}^N, \boldsymbol{w}_R^i)}I(\boldsymbol{x}^N, \boldsymbol{r}^N\vert\boldsymbol{H}[l]) \\
		\text{s.t.}~~\mathbb{E}[\vert\vert \boldsymbol{x} \vert\vert^2_2] \le P_{Tx} \\
		\boldsymbol{w}_R^i = \left[1,e^{j \phi_i},\cdots,e^{j(M_{C} - 1) \phi_i}\right]^H \forall i \in \{1,...,M_{RFE}\},
	\end{gathered}
\end{equation}
with $\boldsymbol{x}^N$ and $\boldsymbol{r}^N$ being $N$ input/output samples of the system. 
Due to the non-linearity of the quantization and the non-trivial problem of finding the optimal beamforming configurations $\boldsymbol{w}_R^i$, we
make a number of approximations that make the expression traceable:
\begin{itemize}
	\item Assume $\boldsymbol{x}(f)$ is Gaussian
	\item For a system with CSIT SVD based precoding is used (SVD of the effective channel after analog combining is used)
	\item $\boldsymbol{w}_R^i$ are selected from the derived finite set separately for each antenna group based on an SNR criteria
	\item Quantization is modeled as additive Gaussian noise with the AQNM model including the off-diagonal elements
\end{itemize}
With this simplifications the $\boldsymbol{w}_R^i$ are already defined and we can transform the problem into a frequency domain equation. 
In \cite{MIMOCAP, Molish2002} the achievable rate of a digital beamforming system without quantization, but considering a multi-path channel is described.
The solution is waterfiling across the frequency bins and the spatial streams. Since for a system with low resolution ADCs the quantization does influence the signal relative to the total power, it
is intuitive to use each frequency bin independent of each other.
Since the optimization is carried out for each frequency bin $f$ separately, the result only is a lower bound to the joint optimization. 
\begin{equation}
	\begin{gathered}
		R \le \int\displaylimits_{f1}^{f2}\max_{\boldsymbol{R}_{\boldsymbol{x}(f)\boldsymbol{x}(f)}}I(\boldsymbol{x}(f), \boldsymbol{r}(f)\vert\boldsymbol{H}^\prime(f))df \\
		\text{s.t.}~~\mathbb{E}[\vert\vert \boldsymbol{x}(f) \vert\vert^2_2] \le P_{Tx} ~\forall f \in [f_1, f_2],
	\end{gathered}
	\label{eq:achievrate}
\end{equation}
with $\boldsymbol{x}(f)$, $\boldsymbol{r}(f)$ and $\boldsymbol{H}^\prime(f)$ being the input/output signal and equivalent channel of frequency bin $f$.
The frequency $f_1$ and $f_2$ mark the borders of the band of interest in the equivalent baseband channel.
If not the whole band covered by the sampling rate is available to the system, the
parameters $f_1$ and $f_2$ have to account for the oversampling. 

Since all signals are represented by Gaussian random variables, we get the following expression for the mutual information:
\begin{equation}
	\begin{gathered}
		 I(\boldsymbol{x}(f), \boldsymbol{r}(f)\vert\boldsymbol{H}^\prime(f)) = \\ 
		\log_2\left(\text{det}\left( \boldsymbol{I} + \boldsymbol{R}^{-1}_{\boldsymbol{\eta}^\prime\boldsymbol{\eta}^\prime} \boldsymbol{H}^\prime(f) \boldsymbol{R}_{\boldsymbol{x}(f)\boldsymbol{x}(f)} \boldsymbol{H}^{\prime H}(f)\right)\right). \\
	\end{gathered}
\end{equation}
For the non-quantized case the result of the optimization is the waterfilling solution. Due to the modeling of the quantization, the effective noise 
covariance matrix $\boldsymbol{R}_{\boldsymbol{\eta}^\prime\boldsymbol{\eta}^\prime}$ and the effective channel 
$ \boldsymbol{H}^\prime(f)$ are dependent on the input covariance matrix $\boldsymbol{R}_{\boldsymbol{x}(f)\boldsymbol{x}(f)}$

In a system without quantization, the covariance $\boldsymbol{R}_{\boldsymbol{x}(f)\boldsymbol{x}(f)}$ would be chosen according to the right singular vectors of $\boldsymbol{H}(f)$
to split the channel in orthogonal subchannels \cite{MIMOCAP}. 
In this scenario $\boldsymbol{R}_{\boldsymbol{x}(f)\boldsymbol{x}(f)}$ would be equal to
\begin{equation}
	\boldsymbol{R}_{\boldsymbol{x}(f)\boldsymbol{x}(f)} = \boldsymbol{V}(f) \boldsymbol{S}(f) \boldsymbol{V}^H(f),
\end{equation}
where $\boldsymbol{V}(f)$ are the eigenvectors of $\boldsymbol{H}^H(f)\boldsymbol{H}(f)$. The diagonal matrix $\boldsymbol{S}(f)$ represents the power allocation to the subchannels. 
The optimal allocation in a system without quantization follows the waterfilling solution. Since $\boldsymbol{R}_{\boldsymbol{\eta}^\prime\boldsymbol{\eta}^\prime}$ and $\boldsymbol{H}^\prime(f)$ actually depend on 
$\boldsymbol{R}_{\boldsymbol{x}(f)\boldsymbol{x}(f)}$, it is difficult to separate the channel into orthogonal subchannels. To make the evaluation traceable, we use the suboptimal precoding vector $\boldsymbol{V}(f)$. 
For the matrix $\boldsymbol{S}(f)$, we test all different possibilities of allocating equal to power to $1$ to $S_{max}$ spatial streams. The number $S_{max}$ is the maximum possible number of spatial streams and is equal to $\text{rank}(\boldsymbol{H}(f))$. 
If we would allow all frequencies to separately allocate the number of streams, we again have a combinatorial problem. Therefore, we check the overall achievable rate for allocating $j$ spatial streams and in the end select one that has the largest achievable
rate.

From Equation \eqref{eq:Rnndash}, we see that
 $\boldsymbol{R}_{\boldsymbol{\eta}^\prime\boldsymbol{\eta}^\prime}$
is not diagonal. In a system, where the noise covariance matrix is known and independent of the transmit covariance one would simply multiply the receive vector with
 $\boldsymbol{R}_{\boldsymbol{\eta}^\prime\boldsymbol{\eta}^\prime}^{-\frac{1}{2}}$. 
This does generate a new system with a different channel $\boldsymbol{R}_{\boldsymbol{\eta}^\prime\boldsymbol{\eta}^\prime}^{-\frac{1}{2}} \boldsymbol{H}^\prime(f)$
 and spatial white noise. Afterwards, the waterfilling solution is applied to the
new channel \cite{MIMOCAP}. In general, the achievable rate increases compared to a system with white noise. In a more abstract way, the reason for the improvement is that 
channels with lower noise power can be used. Dependent on the rank of the channel relative to the number of the receive antennas, the orthogonal subchannels with highest noise power might not be used. 
In a system where the channel and the noise depends on the covariance matrix of the transmit signal, it is very difficult to generate precoding and reception matrices to split the channel into orthogonal subchannels. 
Therefore, with our system, considering the correlation of the quantization noise leads to a decrease in achievable rate.

For both calculation of the achievable rate in Equation \eqref{eq:achievrate} as well as the calculation of the receive signal covariance matrix 
$\boldsymbol{R}_{\boldsymbol{y}_{C}\boldsymbol{y}_{C}}$ in Equation \eqref{eq:Ryycalc}, it is necessary to integrate over the whole signal band from $f_1$ to $f_2$.
Instead of taking infinitely small frequency bins, we approximate the signal band in the interval from $f_1$ to $f_2$ by a finite number of frequency bins. 
We choose the number of frequency bins to make the channel $\boldsymbol{H}(f)$ at each frequency bin
sufficiently flat.
This leads to a good approximation of the achievable rate. Equation \eqref{eq:achievrate} is then reduced to:
\begin{equation}
	\begin{gathered}
		R \le \sum\limits_{f_1}^{f_2}\log_2\left(\text{det}\left( \boldsymbol{I} + \boldsymbol{R}^{-1}_{\boldsymbol{\eta}^\prime\boldsymbol{\eta}^\prime} \boldsymbol{H}^\prime(f) \boldsymbol{R}_{\boldsymbol{x}(f)\boldsymbol{x}(f)} \boldsymbol{H}^{\prime H}(f)\right)\right) \\
		\text{s.t.}~~\mathbb{E}[\vert\vert \boldsymbol{x}(f) \vert\vert^2_2] \le P_{Tx}~\forall f \in [f_1, f_2].
	\end{gathered}
\end{equation}
The receive signal correlation matrix can then be calculated as:
\begin{equation}
	\begin{gathered}
		\boldsymbol{R}_{\boldsymbol{y}_{C}\boldsymbol{y}_{C}} = \\
		\boldsymbol{W}_R^H\left(\sum\limits_{f_1}^{f_2}\boldsymbol{H}(f) \boldsymbol{R}_{\boldsymbol{x}(f)\boldsymbol{x}(f)}\boldsymbol{H}^H(f) + \boldsymbol{R}_{\boldsymbol{\eta}(t)\boldsymbol{\eta}(t)} \right)\boldsymbol{W}_R.
	\end{gathered}
\end{equation}
The channel $\boldsymbol{H}(f)$ or the effective channel $\boldsymbol{H}^\prime(f)$ at the frequency bins $f$ can be calculated from the channel tabs $\boldsymbol{H}[l]$ via the DFT $\mathcal{F}(\cdot)$:
\begin{equation}
	\boldsymbol{H}(f) = \mathcal{F}(\boldsymbol{H}[l]).
\end{equation}

We now have all the necessary mathematical tools to approximate the achievable rate of a multipath channel including quantization effects at the receiver. 
Algorithm \ref{alg:RateQuang} describes our approximation of the achievable rate for these type of systems.

\begin{algorithm}[!tb]
	\caption{Approximation of the achievable rate of a quantized system with noise covariance matrix $\boldsymbol{R}_{\boldsymbol{\eta}\boldsymbol{\eta}}$, multipath channel  $\boldsymbol{H}[l]$ and 
		sum power constraint $P_{Tx}$ and quantization function $Q_b(\cdot)$ with resolution of $b$ bits in the frequency band from $f_1$ to $f_2$.}
	\begin{algorithmic}[1]
    		\Require{$\boldsymbol{R}_{\boldsymbol{\eta}\boldsymbol{\eta}}$, $\boldsymbol{H}[l]$,  $P_{Tx}$, $f_1$, $f_2$ and $Q_b(\cdot)$}
		\State $\rho \gets \mathbb{E}\left[\left\vert a - Q_b(a)\right\vert^2\right]$
    		\State $S_{max} \gets \textrm{rank}\left(\sum\limits^{L-1}_{l=0} \boldsymbol{H}[l]\right)$
		\State $\boldsymbol{H}(f) \gets \mathcal{F}(\boldsymbol{H}[l])$
		\State $[\boldsymbol{V}(f)~\boldsymbol{D}(f)] \gets \textrm{eig}(\boldsymbol{H}^H(f)\boldsymbol{H}(f))~\forall f \in \left[f_1,f_2\right]$
		\For{$j \gets 1 \textrm{ to } S_{max}$}
			\State $\boldsymbol{S} \gets \boldsymbol{0}$
			\State $[\boldsymbol{S}]_{i,i} \gets \frac{P_{Tx}}{j}~\forall i = \{1, \hdots, j\}$
			\State $\boldsymbol{R}_{\boldsymbol{x}(f)\boldsymbol{x}(f)} \gets \boldsymbol{V}(f) \boldsymbol{S} \boldsymbol{V}^H(f)~\forall f \in \left[f_1,f_2\right]$
			\State $\boldsymbol{R}_{\boldsymbol{y}\boldsymbol{y}} \gets \sum\limits_{f_1}^{f_2}\boldsymbol{H}(f) \boldsymbol{R}_{\boldsymbol{x}(f)\boldsymbol{x}(f)}\boldsymbol{H}^H(f) + \boldsymbol{R}_{\boldsymbol{\eta}(t)\boldsymbol{\eta}(t)}$
			\State $\boldsymbol{R}_{\boldsymbol{r}\boldsymbol{r}} \gets T(\boldsymbol{R}_{\boldsymbol{y}\boldsymbol{y}}, Q_b(\cdot))$
			\State $\boldsymbol{R}_{\boldsymbol{e}\boldsymbol{e}} \gets \boldsymbol{R}_{\boldsymbol{r}\boldsymbol{r}} - $
	 		\Statex\hspace{\algorithmicindent}\hspace{\algorithmicindent}{$(1 - \rho)^2 \text{diag}\left( \boldsymbol{R}_{\boldsymbol{y}\boldsymbol{y}}\right)^{-\frac{1}{2}}\boldsymbol{R}_{\boldsymbol{y}\boldsymbol{y}} \text{diag}\left( \boldsymbol{R}_{\boldsymbol{y}\boldsymbol{y}}\right)^{-\frac{1}{2}}$}
			\State $\boldsymbol{F} \gets (1 - \rho) \text{diag}\left( \boldsymbol{R}_{\boldsymbol{y}\boldsymbol{y}}\right)^{-\frac{1}{2}}$
			\State $\boldsymbol{R}_{\boldsymbol{\eta}^\prime\boldsymbol{\eta}^\prime} \gets \boldsymbol{R}_{\boldsymbol{e}\boldsymbol{e}} + \boldsymbol{F} \boldsymbol{R}_{\boldsymbol{\eta}\boldsymbol{\eta}} \boldsymbol{F}^H$
			\State{$\boldsymbol{H}^\prime[l] \gets \boldsymbol{F}\boldsymbol{H}[l]~\forall l \in \{0, \hdots, L-1\}$}
			\State $\boldsymbol{H}^\prime(f) \gets \mathcal{F}(\boldsymbol{H}^\prime[l])$
			\State{$\boldsymbol{A}(f) \gets  \boldsymbol{I} + \boldsymbol{R}^{-1}_{\boldsymbol{\eta}^\prime\boldsymbol{\eta}^\prime} \boldsymbol{H}^\prime(f) \boldsymbol{R}_{\boldsymbol{x}(f)\boldsymbol{x}(f)} \boldsymbol{H}^{\prime H}(f)$}
			\Statex\hspace{\algorithmicindent}\hspace{\algorithmicindent}{$\forall f \in \left[f_1, f_2\right]$}
			\State $R(j) = \sum\limits^{f_2}_{f_1} \log_2\left(\text{det}\left(\boldsymbol{A}(f)\right)\right)$
     	 	\EndFor
		\State $R_{max} \gets \max\limits_j R(j)$
		\State \Return $R_{max}$
  \end{algorithmic}
  \label{alg:RateQuang}
\end{algorithm}

This approximation is modeling a point to point closed loop spatial multiplexing system. There are many different simple modification possible to change the modeled system.
The following are a non-exhaustive list of examples:
\begin{itemize}
	\item Systems without CSIT
	\item Systems with imperfect channel estimation
	\item Systems with multiple terminals communication with base station
	\item Systems with constraint feedback
	\item Systems with multiple terminals and a basestation 
\end{itemize}
Most of these systems can be modeled by changing the constraints on the precoding matrix 
$\boldsymbol{R}_{\boldsymbol{x}(f)\boldsymbol{x}(f)}$ and the channel model.

%% file: AchievableRateComparison/AchievableRateComparison.tex
\section{Evaluation Results}

In this section we evaluate the derived expression for different scenarios. We always include a rate evaluation without quantization. For the system without quantization we apply the waterfilling
solution separate for each frequency bin. For all scenarios the results show the average achievable rate in bps/Hz averaged over 1000 channel realizations. 
\subsection{Comparison to Diagonal Approximation}
This part of the evaluation compares difference in performance when considering the non-diagonal elements in the calculation of $\boldsymbol{R}_{\boldsymbol{r}\boldsymbol{r}}$ and therefore
$\boldsymbol{R}_{\boldsymbol{\eta}^\prime\boldsymbol{\eta}^\prime}$. The model considering only the diagonal elements of $\boldsymbol{R}_{\boldsymbol{r}\boldsymbol{r}}$ was used in
\cite{ONERLOWRES} and \cite{HLOWADC}.

For the evaluation a channel of the first channel model is used. Here $K = 7$ separate paths are received at the same time. Different number of transmit and receive antennas are used.
\ifCLASSOPTIONdraftcls
From Figure \ref{fig:NonDiagVsDiag} (\textbf{A}), 
\else
From Figure \ref{fig:NonDiagVsDiag32x1}, 
\fi
we see that the model considering the off diagonal elements (ND) has a significant lower performance
compared to the model only considering the diagonal elements (D). In fact, for the case of only one transmit antenna ($M_t = 1$) and 1-3 bit A/D conversion, the achievable rate is not
maximized at the highest SNR possible, but rather at a finite SNR between 0 and 10 dB. 
\\ \indent
As discussed in Section \ref{sect:achievablerateExpression},
\ifCLASSOPTIONdraftcls
if we compare the results in Figure \ref{fig:NonDiagVsDiag} (\textbf{A}) 
to the ones in Figure \ref{fig:NonDiagVsDiag} (\textbf{B}). Considering the off diagonal elements
\else
if we compare the results in Figure \ref{fig:NonDiagVsDiag32x1} 
to the ones in Figure \ref{fig:NonDiagVsDiag8x8}. Considering the off diagonal elements
\fi
has only a large influence if the number of receive antennas is larger than the number of transmit antennas. This effect can be explained in the following ways: After spatial whitening, the power distribution of the
effective noise is more non-uniform relative to the system that considers only the diagonal component. 
Since the actual channel and noise covariance matrix depends on the
precoding matrix, it is not possible to decompose the channel into orthogonal subchannels with equal SNR. Thus, we cannot avoid using the channel with high noise variance and therefore the overall performance does
degrade in the quantization noise limited, high SNR regime. This effect is only dominating the performance in the case of high SNR and very low resolution quantization.
The peak in the achievable rate comes from the fact that at a certain SNR the noise provides dithering to randomize this structural performance degradation. 
At the minimum variance noise, where sufficient dithering is provided, is the peak in the performance. This effect is called statistic resonance and can be found in many non-linear systems \cite{Mitaim1998}.
\\ \indent
Another important thing to mention is that in a system with multipath propagation and white noise, the covariance matrix $\boldsymbol{R}_{\boldsymbol{y}\boldsymbol{y}}$ of the receive signal is approximated diagonal. This leads to a
diagonal matrix $\boldsymbol{R}_{\boldsymbol{r}\boldsymbol{r}}$ and therefore spatial white noise of the quantized system.
\ifCLASSOPTIONdraftcls
\begin{figure}
\begin{center}
\normalsize
\begin{tikzpicture}
\begin{groupplot}[group style={group size= 2 by 1, horizontal sep=1.5cm}, width=8cm, height=8cm,
]
    \nextgroupplot[title={(\textbf{A}) $M_R = 32$ and $M_T = 1$}, ylabel={avg. achievable rate [bps/Hz]}, grid,  legend cell align=left,
                  		xlabel={SNR [dB]}, xmin =-30, xmax=30, ymin =0, ymax = 16]
		\addplot[ thick, red, mark=*] table[x = x, y=y]{AchievableRateComparison/PlotData/NonDiagVsDiag/SameTime_32_1_7_CapCSITAvg.txt};
		\addplot[ thick, BTORMIX7] table[x = x, y=y]{AchievableRateComparison/PlotData/NonDiagVsDiag/SameTime_32_1_7_bits_8_CapAQNMDiagCSITAvg.txt};
		\addplot[ thick, BTORMIX7, dashed] table[x = x, y=y]{AchievableRateComparison/PlotData/NonDiagVsDiag/SameTime_32_1_7_bits_8_CapAQNMExactCSITAvg.txt};
		\addplot[ thick, BTORMIX6] table[x = x, y=y]{AchievableRateComparison/PlotData/NonDiagVsDiag/SameTime_32_1_7_bits_7_CapAQNMDiagCSITAvg.txt};
		\addplot[ thick, BTORMIX6, dashed] table[x = x, y=y]{AchievableRateComparison/PlotData/NonDiagVsDiag/SameTime_32_1_7_bits_7_CapAQNMExactCSITAvg.txt};
		\addplot[ thick, BTORMIX5] table[x = x, y=y]{AchievableRateComparison/PlotData/NonDiagVsDiag/SameTime_32_1_7_bits_6_CapAQNMDiagCSITAvg.txt};5
		\addplot[ thick, BTORMIX5, dashed] table[x = x, y=y]{AchievableRateComparison/PlotData/NonDiagVsDiag/SameTime_32_1_7_bits_6_CapAQNMExactCSITAvg.txt};
		\addplot[ thick, BTORMIX4] table[x = x, y=y]{AchievableRateComparison/PlotData/NonDiagVsDiag/SameTime_32_1_7_bits_5_CapAQNMDiagCSITAvg.txt};
		\addplot[ thick, BTORMIX4, dashed] table[x = x, y=y]{AchievableRateComparison/PlotData/NonDiagVsDiag/SameTime_32_1_7_bits_5_CapAQNMExactCSITAvg.txt};
		\addplot[ thick, BTORMIX3] table[x = x, y=y]{AchievableRateComparison/PlotData/NonDiagVsDiag/SameTime_32_1_7_bits_4_CapAQNMDiagCSITAvg.txt};
		\addplot[ thick, BTORMIX3, dashed] table[x = x, y=y]{AchievableRateComparison/PlotData/NonDiagVsDiag/SameTime_32_1_7_bits_4_CapAQNMExactCSITAvg.txt};
		\addplot[ thick, BTORMIX2] table[x = x, y=y]{AchievableRateComparison/PlotData/NonDiagVsDiag/SameTime_32_1_7_bits_3_CapAQNMDiagCSITAvg.txt};
		\addplot[ thick, BTORMIX2, dashed] table[x = x, y=y]{AchievableRateComparison/PlotData/NonDiagVsDiag/SameTime_32_1_7_bits_3_CapAQNMExactCSITAvg.txt};
		\addplot[ thick, BTORMIX1] table[x = x, y=y]{AchievableRateComparison/PlotData/NonDiagVsDiag/SameTime_32_1_7_bits_2_CapAQNMDiagCSITAvg.txt};
		\addplot[ thick, BTORMIX1, dashed] table[x = x, y=y]{AchievableRateComparison/PlotData/NonDiagVsDiag/SameTime_32_1_7_bits_2_CapAQNMExactCSITAvg.txt};
		\addplot[ thick, blue] table[x = x, y=y]{AchievableRateComparison/PlotData/NonDiagVsDiag/SameTime_32_1_7_bits_1_CapAQNMDiagCSITAvg.txt};
		\addplot[ thick, blue, dashed] table[x = x, y=y]{AchievableRateComparison/PlotData/NonDiagVsDiag/SameTime_32_1_7_bits_1_CapAQNMExactCSITAvg.txt};
		\node[anchor=west]  at (axis cs:-7,1){increasing resolution};
   		\node (source) at (axis cs:15,1.5){};  		
   		\node (destination) at (axis cs:20,7){};  		
   		\draw[->, very thick](source)--(destination);
		\coordinate (c1) at (rel axis cs:0,1);
    \nextgroupplot[title={(\textbf{B}) $M_R = 8$ and $M_T = 8$}, grid,  legend cell align=left,
                  		xlabel={SNR [dB]}, xmin =-30, xmax=30, ymin =0, ymax = 50, legend pos=north west,
				legend to name=grouplegend1, legend style={legend columns=7,fill=none,draw=black,anchor=center,align=left, column sep=0.15cm}]
		\addplot[ thick, red, mark=*] table[x = x, y=y]{AchievableRateComparison/PlotData/NonDiagVsDiag/SameTime_8_8_7_CapCSITAvg.txt};
		\addplot[ thick, BTORMIX7] table[x = x, y=y]{AchievableRateComparison/PlotData/NonDiagVsDiag/SameTime_8_8_7_bits_8_CapAQNMDiagCSITAvg.txt};
		\addplot[ thick, BTORMIX7, dashed] table[x = x, y=y]{AchievableRateComparison/PlotData/NonDiagVsDiag/SameTime_8_8_7_bits_8_CapAQNMExactCSITAvg.txt};
		\addplot[ thick, BTORMIX6] table[x = x, y=y]{AchievableRateComparison/PlotData/NonDiagVsDiag/SameTime_8_8_7_bits_7_CapAQNMDiagCSITAvg.txt};
		\addplot[ thick, BTORMIX6, dashed] table[x = x, y=y]{AchievableRateComparison/PlotData/NonDiagVsDiag/SameTime_8_8_7_bits_7_CapAQNMExactCSITAvg.txt};
		\addplot[ thick, BTORMIX5] table[x = x, y=y]{AchievableRateComparison/PlotData/NonDiagVsDiag/SameTime_8_8_7_bits_6_CapAQNMDiagCSITAvg.txt};5
		\addplot[ thick, BTORMIX5, dashed] table[x = x, y=y]{AchievableRateComparison/PlotData/NonDiagVsDiag/SameTime_8_8_7_bits_6_CapAQNMExactCSITAvg.txt};
		\addplot[ thick, BTORMIX4] table[x = x, y=y]{AchievableRateComparison/PlotData/NonDiagVsDiag/SameTime_8_8_7_bits_5_CapAQNMDiagCSITAvg.txt};
		\addplot[ thick, BTORMIX4, dashed] table[x = x, y=y]{AchievableRateComparison/PlotData/NonDiagVsDiag/SameTime_8_8_7_bits_5_CapAQNMExactCSITAvg.txt};
		\addplot[ thick, BTORMIX3] table[x = x, y=y]{AchievableRateComparison/PlotData/NonDiagVsDiag/SameTime_8_8_7_bits_4_CapAQNMDiagCSITAvg.txt};
		\addplot[ thick, BTORMIX3, dashed] table[x = x, y=y]{AchievableRateComparison/PlotData/NonDiagVsDiag/SameTime_8_8_7_bits_4_CapAQNMExactCSITAvg.txt};
		\addplot[ thick, BTORMIX2] table[x = x, y=y]{AchievableRateComparison/PlotData/NonDiagVsDiag/SameTime_8_8_7_bits_3_CapAQNMDiagCSITAvg.txt};
		\addplot[ thick, BTORMIX2, dashed] table[x = x, y=y]{AchievableRateComparison/PlotData/NonDiagVsDiag/SameTime_8_8_7_bits_3_CapAQNMExactCSITAvg.txt};
		\addplot[ thick, BTORMIX1] table[x = x, y=y]{AchievableRateComparison/PlotData/NonDiagVsDiag/SameTime_8_8_7_bits_2_CapAQNMDiagCSITAvg.txt};
		\addplot[ thick, BTORMIX1, dashed] table[x = x, y=y]{AchievableRateComparison/PlotData/NonDiagVsDiag/SameTime_8_8_7_bits_2_CapAQNMExactCSITAvg.txt};
		\addplot[ thick, blue] table[x = x, y=y]{AchievableRateComparison/PlotData/NonDiagVsDiag/SameTime_8_8_7_bits_1_CapAQNMDiagCSITAvg.txt};
		\addplot[ thick, blue, dashed] table[x = x, y=y]{AchievableRateComparison/PlotData/NonDiagVsDiag/SameTime_8_8_7_bits_1_CapAQNMExactCSITAvg.txt};
		\addlegendentry{NQ}
		\addlegendentry{D $8b$}
		\addlegendentry{ND $8b$}
		\addlegendentry{D $7b$}
		\addlegendentry{ND $7b$}
		\addlegendentry{D $6b$}
		\addlegendentry{ND $6b$}
		\addlegendentry{D $5b$}
		\addlegendentry{ND $5b$}
		\addlegendentry{D $4b$}
		\addlegendentry{ND $4b$}
		\addlegendentry{D $3b$}
		\addlegendentry{ND $3b$}
		\addlegendentry{D $2b$}
		\addlegendentry{ND $2b$}
		\addlegendentry{D $1b$}
		\addlegendentry{ND $1b$}
		\coordinate (c2) at (rel axis cs:1,1);
		\node[anchor=west]  at (axis cs:-6,1.2){increasing resolution};
   		\node (source) at (axis cs:15,4){};  		
   		\node (destination) at (axis cs:20,18){};
   		\draw[->, very thick](source)--(destination);
\end{groupplot}
    	\coordinate (c3) at ($(c1)!.5!(c2)$);
    	\node[below] at (c3 |- current bounding box.south)
	{\ref{grouplegend1}};
\end{tikzpicture}
\caption{Achievable rate comparision of digital beamforming with different resolution of the ADC for diagonal (D) and non-diagonal (ND) noise model.}
\label{fig:NonDiagVsDiag}
\end{center}
\end{figure}
\else
\begin{figure}
\begin{center}
\normalsize
\begin{tikzpicture}
    \begin{axis}[width=0.95*8.8cm, height=0.95*8.8cm, ylabel={avg. achievable rate [bps/Hz]}, grid,  legend cell align=left,
                  		xlabel={SNR [dB]}, xmin =-30, xmax=30, ymin =0, ymax = 16, legend pos=north west, legend cell align=left, legend columns=2]
		\addplot[ thick, red, mark=*] table[x = x, y=y]{AchievableRateComparison/PlotData/NonDiagVsDiag/SameTime_32_1_7_CapCSITAvg.txt};
		\addplot[ thick, BTORMIX7] table[x = x, y=y]{AchievableRateComparison/PlotData/NonDiagVsDiag/SameTime_32_1_7_bits_8_CapAQNMDiagCSITAvg.txt};
		\addplot[ thick, BTORMIX7, dashed] table[x = x, y=y]{AchievableRateComparison/PlotData/NonDiagVsDiag/SameTime_32_1_7_bits_8_CapAQNMExactCSITAvg.txt};
		\addplot[ thick, BTORMIX6] table[x = x, y=y]{AchievableRateComparison/PlotData/NonDiagVsDiag/SameTime_32_1_7_bits_7_CapAQNMDiagCSITAvg.txt};
		\addplot[ thick, BTORMIX6, dashed] table[x = x, y=y]{AchievableRateComparison/PlotData/NonDiagVsDiag/SameTime_32_1_7_bits_7_CapAQNMExactCSITAvg.txt};
		\addplot[ thick, BTORMIX5] table[x = x, y=y]{AchievableRateComparison/PlotData/NonDiagVsDiag/SameTime_32_1_7_bits_6_CapAQNMDiagCSITAvg.txt};5
		\addplot[ thick, BTORMIX5, dashed] table[x = x, y=y]{AchievableRateComparison/PlotData/NonDiagVsDiag/SameTime_32_1_7_bits_6_CapAQNMExactCSITAvg.txt};
		\addplot[ thick, BTORMIX4] table[x = x, y=y]{AchievableRateComparison/PlotData/NonDiagVsDiag/SameTime_32_1_7_bits_5_CapAQNMDiagCSITAvg.txt};
		\addplot[ thick, BTORMIX4, dashed] table[x = x, y=y]{AchievableRateComparison/PlotData/NonDiagVsDiag/SameTime_32_1_7_bits_5_CapAQNMExactCSITAvg.txt};
		\addplot[ thick, BTORMIX3] table[x = x, y=y]{AchievableRateComparison/PlotData/NonDiagVsDiag/SameTime_32_1_7_bits_4_CapAQNMDiagCSITAvg.txt};
		\addplot[ thick, BTORMIX3, dashed] table[x = x, y=y]{AchievableRateComparison/PlotData/NonDiagVsDiag/SameTime_32_1_7_bits_4_CapAQNMExactCSITAvg.txt};
		\addplot[ thick, BTORMIX2] table[x = x, y=y]{AchievableRateComparison/PlotData/NonDiagVsDiag/SameTime_32_1_7_bits_3_CapAQNMDiagCSITAvg.txt};
		\addplot[ thick, BTORMIX2, dashed] table[x = x, y=y]{AchievableRateComparison/PlotData/NonDiagVsDiag/SameTime_32_1_7_bits_3_CapAQNMExactCSITAvg.txt};
		\addplot[ thick, BTORMIX1] table[x = x, y=y]{AchievableRateComparison/PlotData/NonDiagVsDiag/SameTime_32_1_7_bits_2_CapAQNMDiagCSITAvg.txt};
		\addplot[ thick, BTORMIX1, dashed] table[x = x, y=y]{AchievableRateComparison/PlotData/NonDiagVsDiag/SameTime_32_1_7_bits_2_CapAQNMExactCSITAvg.txt};
		\addplot[ thick, blue] table[x = x, y=y]{AchievableRateComparison/PlotData/NonDiagVsDiag/SameTime_32_1_7_bits_1_CapAQNMDiagCSITAvg.txt};
		\addplot[ thick, blue, dashed] table[x = x, y=y]{AchievableRateComparison/PlotData/NonDiagVsDiag/SameTime_32_1_7_bits_1_CapAQNMExactCSITAvg.txt};
		\node[anchor=west]  at (axis cs:-7,1){increasing resolution};
		\addlegendentry{NQ}
		\addlegendentry{D $8b$}
		\addlegendentry{ND $8b$}
		\addlegendentry{D $7b$}
		\addlegendentry{ND $7b$}
		\addlegendentry{D $6b$}
		\addlegendentry{ND $6b$}
		\addlegendentry{D $5b$}
		\addlegendentry{ND $5b$}
		\addlegendentry{D $4b$}
		\addlegendentry{ND $4b$}
		\addlegendentry{D $3b$}
		\addlegendentry{ND $3b$}
		\addlegendentry{D $2b$}
		\addlegendentry{ND $2b$}
		\addlegendentry{D $1b$}
		\addlegendentry{ND $1b$}
   		\node (source) at (axis cs:15,1.5){};  		
   		\node (destination) at (axis cs:20,7){};  		
   		\draw[->, very thick](source)--(destination);
		\coordinate (c1) at (rel axis cs:0,1);
     \end{axis}
\end{tikzpicture}
\end{center}
\caption{$M_R = 32$ and $M_T = 1$ digital beamforming with different resolution of the ADC.}
\label{fig:NonDiagVsDiag32x1}
\end{figure}
\begin{figure}
\begin{center}
\normalsize
\begin{tikzpicture}
    \begin{axis}[width=0.95*8.8cm, height=0.95*8.8cm, ylabel={avg. achievable rate [bps/Hz]}, grid,  legend cell align=left,
                  		xlabel={SNR [dB]}, xmin =-30, xmax=30, ymin =0, ymax = 50, legend pos=north west, legend cell align=left, legend columns=2]
		\addplot[ thick, red, mark=*] table[x = x, y=y]{AchievableRateComparison/PlotData/NonDiagVsDiag/SameTime_8_8_7_CapCSITAvg.txt};
		\addplot[ thick, BTORMIX7] table[x = x, y=y]{AchievableRateComparison/PlotData/NonDiagVsDiag/SameTime_8_8_7_bits_8_CapAQNMDiagCSITAvg.txt};
		\addplot[ thick, BTORMIX7, dashed] table[x = x, y=y]{AchievableRateComparison/PlotData/NonDiagVsDiag/SameTime_8_8_7_bits_8_CapAQNMExactCSITAvg.txt};
		\addplot[ thick, BTORMIX6] table[x = x, y=y]{AchievableRateComparison/PlotData/NonDiagVsDiag/SameTime_8_8_7_bits_7_CapAQNMDiagCSITAvg.txt};
		\addplot[ thick, BTORMIX6, dashed] table[x = x, y=y]{AchievableRateComparison/PlotData/NonDiagVsDiag/SameTime_8_8_7_bits_7_CapAQNMExactCSITAvg.txt};
		\addplot[ thick, BTORMIX5] table[x = x, y=y]{AchievableRateComparison/PlotData/NonDiagVsDiag/SameTime_8_8_7_bits_6_CapAQNMDiagCSITAvg.txt};5
		\addplot[ thick, BTORMIX5, dashed] table[x = x, y=y]{AchievableRateComparison/PlotData/NonDiagVsDiag/SameTime_8_8_7_bits_6_CapAQNMExactCSITAvg.txt};
		\addplot[ thick, BTORMIX4] table[x = x, y=y]{AchievableRateComparison/PlotData/NonDiagVsDiag/SameTime_8_8_7_bits_5_CapAQNMDiagCSITAvg.txt};
		\addplot[ thick, BTORMIX4, dashed] table[x = x, y=y]{AchievableRateComparison/PlotData/NonDiagVsDiag/SameTime_8_8_7_bits_5_CapAQNMExactCSITAvg.txt};
		\addplot[ thick, BTORMIX3] table[x = x, y=y]{AchievableRateComparison/PlotData/NonDiagVsDiag/SameTime_8_8_7_bits_4_CapAQNMDiagCSITAvg.txt};
		\addplot[ thick, BTORMIX3, dashed] table[x = x, y=y]{AchievableRateComparison/PlotData/NonDiagVsDiag/SameTime_8_8_7_bits_4_CapAQNMExactCSITAvg.txt};
		\addplot[ thick, BTORMIX2] table[x = x, y=y]{AchievableRateComparison/PlotData/NonDiagVsDiag/SameTime_8_8_7_bits_3_CapAQNMDiagCSITAvg.txt};
		\addplot[ thick, BTORMIX2, dashed] table[x = x, y=y]{AchievableRateComparison/PlotData/NonDiagVsDiag/SameTime_8_8_7_bits_3_CapAQNMExactCSITAvg.txt};
		\addplot[ thick, BTORMIX1] table[x = x, y=y]{AchievableRateComparison/PlotData/NonDiagVsDiag/SameTime_8_8_7_bits_2_CapAQNMDiagCSITAvg.txt};
		\addplot[ thick, BTORMIX1, dashed] table[x = x, y=y]{AchievableRateComparison/PlotData/NonDiagVsDiag/SameTime_8_8_7_bits_2_CapAQNMExactCSITAvg.txt};
		\addplot[ thick, blue] table[x = x, y=y]{AchievableRateComparison/PlotData/NonDiagVsDiag/SameTime_8_8_7_bits_1_CapAQNMDiagCSITAvg.txt};
		\addplot[ thick, blue, dashed] table[x = x, y=y]{AchievableRateComparison/PlotData/NonDiagVsDiag/SameTime_8_8_7_bits_1_CapAQNMExactCSITAvg.txt};
		\addlegendentry{NQ}
		\addlegendentry{D $8b$}
		\addlegendentry{ND $8b$}
		\addlegendentry{D $7b$}
		\addlegendentry{ND $7b$}
		\addlegendentry{D $6b$}
		\addlegendentry{ND $6b$}
		\addlegendentry{D $5b$}
		\addlegendentry{ND $5b$}
		\addlegendentry{D $4b$}
		\addlegendentry{ND $4b$}
		\addlegendentry{D $3b$}
		\addlegendentry{ND $3b$}
		\addlegendentry{D $2b$}
		\addlegendentry{ND $2b$}
		\addlegendentry{D $1b$}
		\addlegendentry{ND $1b$}
		\coordinate (c2) at (rel axis cs:1,1);
		\node[anchor=west]  at (axis cs:-6,1.2){increasing resolution};
   		\node (source) at (axis cs:15,4){};  		
   		\node (destination) at (axis cs:20,18){};
   		\draw[->, very thick](source)--(destination);
     \end{axis}
\end{tikzpicture}
\end{center}
\caption{$M_R = 8$ and $M_T = 8$ digital beamforming with different resolution of the ADC.}
\label{fig:NonDiagVsDiag8x8}
\end{figure}
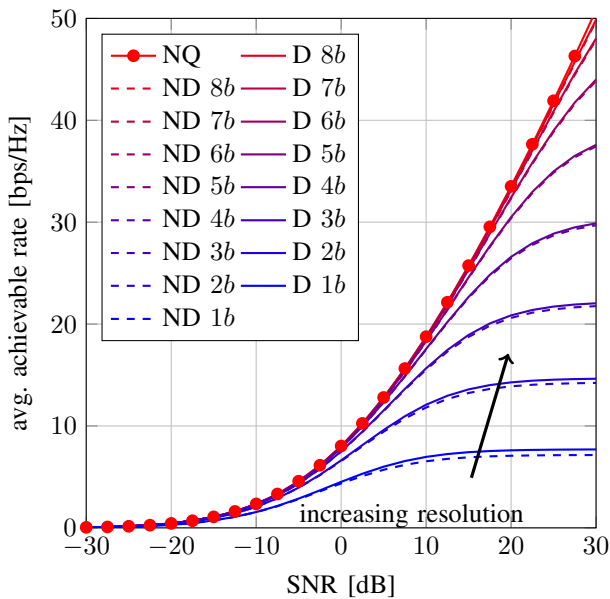
\fi
\subsection{Influence of AGC Imperfection}
In this evaluation, we show the influence of AGC imperfections on the performance. To simplify the evaluation, we choose a SISO system with the 
simple multipath model described in the signal model with the parameters $L = 32$, $P = 16$ and $\beta = 0.35$. For an imperfect AGC, the assumption that the receive signal $\boldsymbol{r}[n]$
and the quantization error $\boldsymbol{e}[n]$ are independent is no longer satisfied. Since all our formulas for modeling the quantization are based on this assumption 
$\mathbb{E}\left[r_i[n] e_i^H[n]\right] =0~\forall i = \left\{1, \hdots, M_{R}\right\}$, they are no longer valid in the case of an imperfect adapted AGC. 
We can enforce this orthogonality by scaling the signal $\boldsymbol{r}[n]$. The scaling factor $\zeta$ is equal to:
\begin{equation}
	\zeta = \frac{\mathbb{E}\left[a Q_b(a)\right]}{\mathbb{E}\left[Q_b(a)^2\right]},
\end{equation} 
with $a$ being a real Gaussian random variable with unit variance and zero mean. After this scaling we can use the derived formulas as before. 

\begin{figure}
\begin{center}
\begin{tikzpicture}{avg. achievable rate [bps/Hz]}
    \begin{axis}[width=0.95*8.8cm, height=0.95*8.8cm, ylabel={avg. achievable rate [bps/Hz]}, grid,  legend cell align=left,
                  		xlabel={SNR [dB]}, xmin =-30, xmax=30, ymin =0, ymax = 3, legend pos=north west, legend cell align=left]
		\addplot[ thick, red, mark=*] table[x = x, y=y]{AchievableRateComparison/PlotData/AGCError/FullSNR/MPath_1_1_Offset_bits_1_Offset_0_CapCSITAvg.txt};
		\addplot[ thick, green] table[x = x, y=y]{AchievableRateComparison/PlotData/AGCError/FullSNR/MPath_1_1_Offset_bits_2_Offset_0_CapAQNMExactCSITAvg.txt};
		\addplot[ thick, GTOBMIX1, mark=o] table[x = x, y=y]{AchievableRateComparison/PlotData/AGCError/FullSNR/MPath_1_1_Offset_bits_2_Offset_10_CapAQNMExactCSITAvg.txt};
		\addplot[ thick, GTOBMIX1, mark=|] table[x = x, y=y]{AchievableRateComparison/PlotData/AGCError/FullSNR/MPath_1_1_Offset_bits_2_Offset_-10_CapAQNMExactCSITAvg.txt};
		\addplot[ thick, GTOBMIX2, mark=square] table[x = x, y=y]{AchievableRateComparison/PlotData/AGCError/FullSNR/MPath_1_1_Offset_bits_2_Offset_20_CapAQNMExactCSITAvg.txt};
		\addplot[ thick, GTOBMIX2, mark=star] table[x = x, y=y]{AchievableRateComparison/PlotData/AGCError/FullSNR/MPath_1_1_Offset_bits_2_Offset_-20_CapAQNMExactCSITAvg.txt};
		\addplot[ thick, GTOBMIX3, mark=triangle] table[x = x, y=y]{AchievableRateComparison/PlotData/AGCError/FullSNR/MPath_1_1_Offset_bits_2_Offset_40_CapAQNMExactCSITAvg.txt};
		\addplot[ thick, GTOBMIX3, mark=x] table[x = x, y=y]{AchievableRateComparison/PlotData/AGCError/FullSNR/MPath_1_1_Offset_bits_2_Offset_-40_CapAQNMExactCSITAvg.txt};
		\addplot[ thick, GTOBMIX4, mark=diamond] table[x = x, y=y]{AchievableRateComparison/PlotData/AGCError/FullSNR/MPath_1_1_Offset_bits_2_Offset_80_CapAQNMExactCSITAvg.txt};
		\addplot[ thick, GTOBMIX4, mark=Mercedes star flipped] table[x = x, y=y]{AchievableRateComparison/PlotData/AGCError/FullSNR/MPath_1_1_Offset_bits_2_Offset_-80_CapAQNMExactCSITAvg.txt};
		\addplot[ thick, blue, dashed] table[x = x, y=y]{AchievableRateComparison/PlotData/AGCError/FullSNR/MPath_1_1_Offset_bits_1_Offset_0_CapAQNMExactCSITAvg.txt};
		\addlegendentry{NQ}
		\addlegendentry{$2b$}
		\addlegendentry{$2b$ 10\%}
		\addlegendentry{$2b$ -10\%}
		\addlegendentry{$2b$ 20\%}
		\addlegendentry{$2b$ -20\%}
		\addlegendentry{$2b$ 40\%}
		\addlegendentry{$2b$ -40\%}
		\addlegendentry{$2b$ 80\%}
		\addlegendentry{$2b$ -80\%}
		\addlegendentry{$1b$}
     \end{axis}
\end{tikzpicture}
\end{center}
\caption{SISO system with imperfect AGC and 2 bit ADC resolution.}
\label{fig:AGC2bfullSNR}
\end{figure}
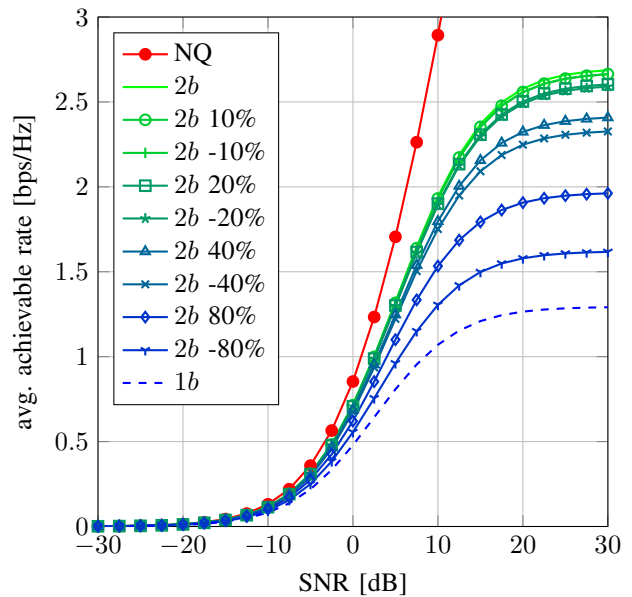

The graphs in Figure \ref{fig:AGC2bfullSNR}
show the average achievable rate with 2 bit resolution and different
offset relative to the optimal power at the VGA output. 
The power after the VGA is defined as
\begin{equation}
	\Omega = \Omega_{mq} (1 - \epsilon_{\text{AGC}}),
\end{equation}
where $\Omega_{mq}$ and $\epsilon_{\text{AGC}}$ are the signal variance resulting in the minimal distortion and the AGC error. 

The graphs shows that an error in the range of -20\% to 20\% has only a minor impact on the performance. But as soon as the error is 
larger than 20\%, the performance decreases dramatically. Ultimately, the quantization converges to 1-bit quantization and therefore also our achievable rate converges to the one of 1-bit quantization.
We can also observe the performance penalty for a larger negative or positive error is different. 

\subsection{Downlink (DL) Point to Point Scenario}
\ifCLASSOPTIONdraftcls
\begin{figure}
\begin{center}
\begin{tikzpicture}
\begin{groupplot}[group style={group size= 2 by 1, horizontal sep=1.5cm}, width=8cm, height=8cm]
    \nextgroupplot[title={(\textbf{A}) $M_R = 8$ and $M_T = 64$}, ylabel={avg. achievable rate [bps/Hz]}, grid,  legend cell align=left,
                  		xlabel={SNR [dB]}, xmin =-30, xmax=30, ymin =0, ymax = 80]
		\addplot[ thick, red, mark=*] table[x = x, y=y]{AchievableRateComparison/PlotData/DL/SE/Multipath_8_64_16_32_3.5e-01_1000_CapCSITAvg.txt};
		\addplot[ thick, BTORMIX7] table[x = x, y=y]{AchievableRateComparison/PlotData/DL/SE/Multipath_8_64_16_32_3.5e-01_1000_bits_8_CapAQNMExactCSITAvg.txt};
		\addplot[ thick, BTORMIX6] table[x = x, y=y]{AchievableRateComparison/PlotData/DL/SE/Multipath_8_64_16_32_3.5e-01_1000_bits_7_CapAQNMExactCSITAvg.txt};
		\addplot[ thick, BTORMIX5] table[x = x, y=y]{AchievableRateComparison/PlotData/DL/SE/Multipath_8_64_16_32_3.5e-01_1000_bits_6_CapAQNMExactCSITAvg.txt};
		\addplot[ thick, BTORMIX4] table[x = x, y=y]{AchievableRateComparison/PlotData/DL/SE/Multipath_8_64_16_32_3.5e-01_1000_bits_5_CapAQNMExactCSITAvg.txt};
		\addplot[ thick, BTORMIX3] table[x = x, y=y]{AchievableRateComparison/PlotData/DL/SE/Multipath_8_64_16_32_3.5e-01_1000_bits_4_CapAQNMExactCSITAvg.txt};
		\addplot[ thick, BTORMIX2] table[x = x, y=y]{AchievableRateComparison/PlotData/DL/SE/Multipath_8_64_16_32_3.5e-01_1000_bits_3_CapAQNMExactCSITAvg.txt};
		\addplot[ thick, BTORMIX1] table[x = x, y=y]{AchievableRateComparison/PlotData/DL/SE/Multipath_8_64_16_32_3.5e-01_1000_bits_2_CapAQNMExactCSITAvg.txt};
		\addplot[ thick, blue] table[x = x, y=y]{AchievableRateComparison/PlotData/DL/SE/Multipath_8_64_16_32_3.5e-01_1000_bits_1_CapAQNMExactCSITAvg.txt};
		\addplot[ thick, green, mark=triangle*, mark options={solid, fill=green}] table[x = x, y=y]{AchievableRateComparison/PlotData/DL/SE/Multipath_8_64_16_32_3.5e-01_1000_MAC_4_CapHBFNQAvgCombined.txt};
		\addplot[ thick, dashed, BTOGMIX7] table[x = x, y=y]{AchievableRateComparison/PlotData/DL/SE/Multipath_8_64_16_32_3.5e-01_1000_bits_8_MAC_4_CapHBFAQNMExactCSITAvgCombined.txt};
		\addplot[ thick, dashed, BTOGMIX6] table[x = x, y=y]{AchievableRateComparison/PlotData/DL/SE/Multipath_8_64_16_32_3.5e-01_1000_bits_7_MAC_4_CapHBFAQNMExactCSITAvgCombined.txt};
		\addplot[ thick, dashed, BTOGMIX5] table[x = x, y=y]{AchievableRateComparison/PlotData/DL/SE/Multipath_8_64_16_32_3.5e-01_1000_bits_6_MAC_4_CapHBFAQNMExactCSITAvgCombined.txt};
		\addplot[ thick, dashed, BTOGMIX4] table[x = x, y=y]{AchievableRateComparison/PlotData/DL/SE/Multipath_8_64_16_32_3.5e-01_1000_bits_5_MAC_4_CapHBFAQNMExactCSITAvgCombined.txt};
		\addplot[ thick, dashed, BTOGMIX3] table[x = x, y=y]{AchievableRateComparison/PlotData/DL/SE/Multipath_8_64_16_32_3.5e-01_1000_bits_4_MAC_4_CapHBFAQNMExactCSITAvgCombined.txt};
		\addplot[ thick, dashed, BTOGMIX2] table[x = x, y=y]{AchievableRateComparison/PlotData/DL/SE/Multipath_8_64_16_32_3.5e-01_1000_bits_3_MAC_4_CapHBFAQNMExactCSITAvgCombined.txt};
		\addplot[ thick, dashed, BTOGMIX1] table[x = x, y=y]{AchievableRateComparison/PlotData/DL/SE/Multipath_8_64_16_32_3.5e-01_1000_bits_2_MAC_4_CapHBFAQNMExactCSITAvgCombined.txt};
		\addplot[ thick, dashed, blue] table[x = x, y=y]{AchievableRateComparison/PlotData/DL/SE/Multipath_8_64_16_32_3.5e-01_1000_bits_1_MAC_4_CapHBFAQNMExactCSITAvgCombined.txt};
		\coordinate (c1) at (rel axis cs:0,1);
    \nextgroupplot[title={(\textbf{B}) $M_R = 64$ and $M_T = 8$}, grid,  legend cell align=left,
                  		xlabel={SNR [dB]}, xmin =-30, xmax=30, ymin =0, ymax = 80, legend pos=north west,
				legend to name=grouplegend2, legend style={legend columns=6,fill=none,draw=black,anchor=center,align=left, column sep=0.15cm}]
		\addplot[ thick, red, mark=*] table[x = x, y=y]{AchievableRateComparison/PlotData/UL/SE/Multipath_64_8_16_32_3.5e-01_1000_CapCSITAvg.txt};
		\addplot[ thick, BTORMIX7] table[x = x, y=y]{AchievableRateComparison/PlotData/UL/SE/Multipath_64_8_16_32_3.5e-01_1000_bits_8_CapAQNMExactCSITAvg.txt};
		\addplot[ thick, BTORMIX6] table[x = x, y=y]{AchievableRateComparison/PlotData/UL/SE/Multipath_64_8_16_32_3.5e-01_1000_bits_7_CapAQNMExactCSITAvg.txt};
		\addplot[ thick, BTORMIX5] table[x = x, y=y]{AchievableRateComparison/PlotData/UL/SE/Multipath_64_8_16_32_3.5e-01_1000_bits_6_CapAQNMExactCSITAvg.txt};
		\addplot[ thick, BTORMIX4] table[x = x, y=y]{AchievableRateComparison/PlotData/UL/SE/Multipath_64_8_16_32_3.5e-01_1000_bits_5_CapAQNMExactCSITAvg.txt};
		\addplot[ thick, BTORMIX3] table[x = x, y=y]{AchievableRateComparison/PlotData/UL/SE/Multipath_64_8_16_32_3.5e-01_1000_bits_4_CapAQNMExactCSITAvg.txt};
		\addplot[ thick, BTORMIX2] table[x = x, y=y]{AchievableRateComparison/PlotData/UL/SE/Multipath_64_8_16_32_3.5e-01_1000_bits_3_CapAQNMExactCSITAvg.txt};
		\addplot[ thick, BTORMIX1] table[x = x, y=y]{AchievableRateComparison/PlotData/UL/SE/Multipath_64_8_16_32_3.5e-01_1000_bits_2_CapAQNMExactCSITAvg.txt};
		\addplot[ thick, blue] table[x = x, y=y]{AchievableRateComparison/PlotData/UL/SE/Multipath_64_8_16_32_3.5e-01_1000_bits_1_CapAQNMExactCSITAvg.txt};
		\addplot[ thick, green, mark=triangle*, mark options={solid, fill=green}] table[x = x, y=y]{AchievableRateComparison/PlotData/UL/SE/Multipath_64_8_16_32_3.5e-01_1000_MAC_4_CapHBFNQAvgCombined.txt};
		\addplot[ thick, dashed, BTOGMIX7] table[x = x, y=y]{AchievableRateComparison/PlotData/UL/SE/Multipath_64_8_16_32_3.5e-01_1000_bits_8_MAC_4_CapHBFAQNMExactCSITAvgCombined.txt};
		\addplot[ thick, dashed, BTOGMIX6] table[x = x, y=y]{AchievableRateComparison/PlotData/UL/SE/Multipath_64_8_16_32_3.5e-01_1000_bits_7_MAC_4_CapHBFAQNMExactCSITAvgCombined.txt};
		\addplot[ thick, dashed, BTOGMIX5] table[x = x, y=y]{AchievableRateComparison/PlotData/UL/SE/Multipath_64_8_16_32_3.5e-01_1000_bits_6_MAC_4_CapHBFAQNMExactCSITAvgCombined.txt};
		\addplot[ thick, dashed, BTOGMIX4] table[x = x, y=y]{AchievableRateComparison/PlotData/UL/SE/Multipath_64_8_16_32_3.5e-01_1000_bits_5_MAC_4_CapHBFAQNMExactCSITAvgCombined.txt};
		\addplot[ thick, dashed, BTOGMIX3] table[x = x, y=y]{AchievableRateComparison/PlotData/UL/SE/Multipath_64_8_16_32_3.5e-01_1000_bits_4_MAC_4_CapHBFAQNMExactCSITAvgCombined.txt};
		\addplot[ thick, dashed, BTOGMIX2] table[x = x, y=y]{AchievableRateComparison/PlotData/UL/SE/Multipath_64_8_16_32_3.5e-01_1000_bits_3_MAC_4_CapHBFAQNMExactCSITAvgCombined.txt};
		\addplot[ thick, dashed, BTOGMIX1] table[x = x, y=y]{AchievableRateComparison/PlotData/UL/SE/Multipath_64_8_16_32_3.5e-01_1000_bits_2_MAC_4_CapHBFAQNMExactCSITAvgCombined.txt};
		\addplot[ thick, dashed, blue] table[x = x, y=y]{AchievableRateComparison/PlotData/UL/SE/Multipath_64_8_16_32_3.5e-01_1000_bits_1_MAC_4_CapHBFAQNMExactCSITAvgCombined.txt};
		\addlegendentry{DBF}
		\addlegendentry{DBF 8$b$}
		\addlegendentry{DBF 7$b$}
		\addlegendentry{DBF 6$b$}
		\addlegendentry{DBF 5$b$}
		\addlegendentry{DBF 4$b$}
		\addlegendentry{DBF 3$b$}
		\addlegendentry{DBF 2$b$}
		\addlegendentry{DBF 1$b$}
		\addlegendentry{HBF}
		\addlegendentry{HBF 8$b$}
		\addlegendentry{HBF 7$b$}
		\addlegendentry{HBF 6$b$}
		\addlegendentry{HBF 5$b$}
		\addlegendentry{HBF 4$b$}
		\addlegendentry{HBF 3$b$}
		\addlegendentry{HBF 2$b$}
		\addlegendentry{HBF 1$b$}
		\coordinate (c2) at (rel axis cs:1,1);
	\end{groupplot}
    	\coordinate (c3) at ($(c1)!.5!(c2)$);
    	\node[below] at (c3 |- current bounding box.south)
	{\ref{grouplegend2}};
\end{tikzpicture}
\end{center}
\caption{Results with $M_{C} = 4$ and different resolution of the ADC $b$ and different antenna configuration.}
\label{fig:DLULMACSE}
\end{figure}
\else
\begin{figure}
\begin{center}
\begin{tikzpicture}
    \begin{axis}[width=0.95*8.8cm, height=0.95*8.8cm, ylabel={avg. achievable rate [bps/Hz]}, grid,  legend cell align=left,
                  		xlabel={SNR [dB]}, xmin =-30, xmax=30, ymin =0, ymax = 80, legend pos=north west, legend cell align=left, legend columns=2]
		\addplot[ thick, red, mark=*] table[x = x, y=y]{AchievableRateComparison/PlotData/DL/SE/Multipath_8_64_16_32_3.5e-01_1000_CapCSITAvg.txt};
		\addplot[ thick, green, mark=triangle*, mark options={solid, fill=green}] table[x = x, y=y]{AchievableRateComparison/PlotData/DL/SE/Multipath_8_64_16_32_3.5e-01_1000_MAC_4_CapHBFNQAvgCombined.txt};
		\addplot[ thick, BTORMIX7] table[x = x, y=y]{AchievableRateComparison/PlotData/DL/SE/Multipath_8_64_16_32_3.5e-01_1000_bits_8_CapAQNMExactCSITAvg.txt};
		\addplot[ thick, dashed, BTOGMIX7] table[x = x, y=y]{AchievableRateComparison/PlotData/DL/SE/Multipath_8_64_16_32_3.5e-01_1000_bits_8_MAC_4_CapHBFAQNMExactCSITAvgCombined.txt};
		\addplot[ thick, BTORMIX6] table[x = x, y=y]{AchievableRateComparison/PlotData/DL/SE/Multipath_8_64_16_32_3.5e-01_1000_bits_7_CapAQNMExactCSITAvg.txt};
		\addplot[ thick, dashed, BTOGMIX6] table[x = x, y=y]{AchievableRateComparison/PlotData/DL/SE/Multipath_8_64_16_32_3.5e-01_1000_bits_7_MAC_4_CapHBFAQNMExactCSITAvgCombined.txt};
		\addplot[ thick, BTORMIX5] table[x = x, y=y]{AchievableRateComparison/PlotData/DL/SE/Multipath_8_64_16_32_3.5e-01_1000_bits_6_CapAQNMExactCSITAvg.txt};
		\addplot[ thick, dashed, BTOGMIX5] table[x = x, y=y]{AchievableRateComparison/PlotData/DL/SE/Multipath_8_64_16_32_3.5e-01_1000_bits_6_MAC_4_CapHBFAQNMExactCSITAvgCombined.txt};
		\addplot[ thick, BTORMIX4] table[x = x, y=y]{AchievableRateComparison/PlotData/DL/SE/Multipath_8_64_16_32_3.5e-01_1000_bits_5_CapAQNMExactCSITAvg.txt};
		\addplot[ thick, dashed, BTOGMIX4] table[x = x, y=y]{AchievableRateComparison/PlotData/DL/SE/Multipath_8_64_16_32_3.5e-01_1000_bits_5_MAC_4_CapHBFAQNMExactCSITAvgCombined.txt};
		\addplot[ thick, BTORMIX3] table[x = x, y=y]{AchievableRateComparison/PlotData/DL/SE/Multipath_8_64_16_32_3.5e-01_1000_bits_4_CapAQNMExactCSITAvg.txt};
		\addplot[ thick, dashed, BTOGMIX3] table[x = x, y=y]{AchievableRateComparison/PlotData/DL/SE/Multipath_8_64_16_32_3.5e-01_1000_bits_4_MAC_4_CapHBFAQNMExactCSITAvgCombined.txt};
		\addplot[ thick, BTORMIX2] table[x = x, y=y]{AchievableRateComparison/PlotData/DL/SE/Multipath_8_64_16_32_3.5e-01_1000_bits_3_CapAQNMExactCSITAvg.txt};
		\addplot[ thick, dashed, BTOGMIX2] table[x = x, y=y]{AchievableRateComparison/PlotData/DL/SE/Multipath_8_64_16_32_3.5e-01_1000_bits_3_MAC_4_CapHBFAQNMExactCSITAvgCombined.txt};
		\addplot[ thick, BTORMIX1] table[x = x, y=y]{AchievableRateComparison/PlotData/DL/SE/Multipath_8_64_16_32_3.5e-01_1000_bits_2_CapAQNMExactCSITAvg.txt};
		\addplot[ thick, dashed, BTOGMIX1] table[x = x, y=y]{AchievableRateComparison/PlotData/DL/SE/Multipath_8_64_16_32_3.5e-01_1000_bits_2_MAC_4_CapHBFAQNMExactCSITAvgCombined.txt};
		\addplot[ thick, blue] table[x = x, y=y]{AchievableRateComparison/PlotData/DL/SE/Multipath_8_64_16_32_3.5e-01_1000_bits_1_CapAQNMExactCSITAvg.txt};
		\addplot[ thick, dashed, blue] table[x = x, y=y]{AchievableRateComparison/PlotData/DL/SE/Multipath_8_64_16_32_3.5e-01_1000_bits_1_MAC_4_CapHBFAQNMExactCSITAvgCombined.txt};

		\addlegendentry{DBF}
		\addlegendentry{HBF}
		\addlegendentry{DBF 8$b$}
		\addlegendentry{HBF 8$b$}
		\addlegendentry{DBF 7$b$}
		\addlegendentry{HBF 7$b$}
		\addlegendentry{DBF 6$b$}
		\addlegendentry{HBF 6$b$}
		\addlegendentry{DBF 5$b$}
		\addlegendentry{HBF 5$b$}
		\addlegendentry{DBF 4$b$}
		\addlegendentry{HBF 4$b$}
		\addlegendentry{DBF 3$b$}
		\addlegendentry{HBF 3$b$}
		\addlegendentry{DBF 2$b$}
		\addlegendentry{HBF 2$b$}
		\addlegendentry{DBF 1$b$}
		\addlegendentry{HBF 1$b$}
     \end{axis}
\end{tikzpicture}
\end{center}
\caption{$M_R = 8$ and $M_T = 64$ $M_{C} = 4$ different resolution of the ADC $b$.}
\label{fig:DLMAC4SE}
\end{figure}
\fi
\ifCLASSOPTIONdraftcls
\begin{figure}
\begin{center}
\begin{tikzpicture}
\begin{groupplot}[group style={group size= 2 by 1, horizontal sep=1.5cm}, width=8cm, height=8cm]
    \nextgroupplot[title={(\textbf{A}) $\text{SNR} = -15\text{dB}$}, xlabel={avg. achievable rate [bps/Hz]}, grid, xmin = 1.1, xmax = 4, ymin=10, ymax =29,
                  		ylabel={energy efficiency [bps/J]}, legend pos=north west, legend cell align=left]
		\addplot[ thick, red, mark=*] table[x = y, y=x]{AchievableRateComparison/PlotData/DL/EE/Multipath_8_64_16_32_3.5e-01_1000_SNR_-150_EE_DBF.txt};
		\addplot[ thick, green, mark=triangle*] table[x = y, y=x]{AchievableRateComparison/PlotData/DL/EE/Multipath_8_64_16_32_3.5e-01_1000_MAC_2_SNR_-150_EE_HBF.txt};
		\addplot[ thick, green, mark=square*] table[x = y, y=x]{AchievableRateComparison/PlotData/DL/EE/Multipath_8_64_16_32_3.5e-01_1000_MAC_4_SNR_-150_EE_HBF.txt};
		\addplot[ thick, green, mark=diamond*] table[x = y, y=x]{AchievableRateComparison/PlotData/DL/EE/Multipath_8_64_16_32_3.5e-01_1000_MAC_8_SNR_-150_EE_HBF.txt};
		\node[anchor=west]  at (axis cs:1.5,11){increasing resolution};
   		\node (source) at (axis cs:1.5,11.5){};  		
   		\node (destination) at (axis cs:2,16){};
   		\draw[->, very thick](source)--(destination);
		\coordinate (c1) at (rel axis cs:0,1);
    \nextgroupplot[title={(\textbf{B}) $\text{SNR} = 0\text{dB}$}, xlabel={avg. achievable rate [bps/Hz]}, grid, xmin = 1, xmax = 16.2, ymin=10, ymax =115,
				legend to name=grouplegend3, legend style={legend columns=4,fill=none,draw=black,anchor=center,align=left, column sep=0.15cm}]
		\addplot[ thick, red, mark=*] table[x = y, y=x]{AchievableRateComparison/PlotData/DL/EE/Multipath_8_64_16_32_3.5e-01_1000_SNR_0_EE_DBF.txt};
		\addplot[ thick, green, mark=triangle*] table[x = y, y=x]{AchievableRateComparison/PlotData/DL/EE/Multipath_8_64_16_32_3.5e-01_1000_MAC_2_SNR_0_EE_HBF.txt};
		\addplot[ thick, green, mark=square*] table[x = y, y=x]{AchievableRateComparison/PlotData/DL/EE/Multipath_8_64_16_32_3.5e-01_1000_MAC_4_SNR_0_EE_HBF.txt};
		\addplot[ thick, green, mark=diamond*] table[x = y, y=x]{AchievableRateComparison/PlotData/DL/EE/Multipath_8_64_16_32_3.5e-01_1000_MAC_8_SNR_0_EE_HBF.txt};
		\addlegendentry{DBF}
		\addlegendentry{HBF $M_{RFE}~4$}
		\addlegendentry{HBF $M_{RFE}~2$}
		\addlegendentry{HBF $M_{RFE}~1$}
		\node[anchor=west]  at (axis cs:2.5,15){increasing resolution};
   		\node (source) at (axis cs:2.5,17){};  		
   		\node (destination) at (axis cs:5,35){};
   		\draw[->, very thick](source)--(destination);
		\coordinate (c2) at (rel axis cs:1,1);
	\end{groupplot}
    	\coordinate (c3) at ($(c1)!.5!(c2)$);
    	\node[below] at (c3 |- current bounding box.south)
	{\ref{grouplegend3}};
\end{tikzpicture}
\end{center}
\caption{$M_R = 8$ and $M_T = 64$ different resolution of the ADC $b$ and different SNR.}
\label{fig:DLEE}
\end{figure}
\else
\begin{figure}
\begin{center}
\begin{tikzpicture}
    \begin{axis}[width=0.95*8.8cm, height=0.95*8.8cm, xlabel={avg. achievable rate [bps/Hz]}, grid, xmin = 1.1, xmax = 4, ymin=10, ymax =29,
                  		ylabel={energy efficiency [bps/J]}, legend pos=north west, legend cell align=left]
		\addplot[ thick, red, mark=*] table[x = y, y=x]{AchievableRateComparison/PlotData/DL/EE/Multipath_8_64_16_32_3.5e-01_1000_SNR_-150_EE_DBF.txt};
		\addplot[ thick, green, mark=triangle*] table[x = y, y=x]{AchievableRateComparison/PlotData/DL/EE/Multipath_8_64_16_32_3.5e-01_1000_MAC_2_SNR_-150_EE_HBF.txt};
		\addplot[ thick, green, mark=square*] table[x = y, y=x]{AchievableRateComparison/PlotData/DL/EE/Multipath_8_64_16_32_3.5e-01_1000_MAC_4_SNR_-150_EE_HBF.txt};
		\addplot[ thick, green, mark=diamond*] table[x = y, y=x]{AchievableRateComparison/PlotData/DL/EE/Multipath_8_64_16_32_3.5e-01_1000_MAC_8_SNR_-150_EE_HBF.txt};
		\addlegendentry{DBF}
		\addlegendentry{HBF $M_{RFE}~4$}
		\addlegendentry{HBF $M_{RFE}~2$}
		\addlegendentry{HBF $M_{RFE}~1$}
		\node[anchor=west]  at (axis cs:1.5,11){increasing resolution};
   		\node (source) at (axis cs:1.5,11.5){};  		
   		\node (destination) at (axis cs:2,16){};
   		\draw[->, very thick](source)--(destination);
     \end{axis}
\end{tikzpicture}
\end{center}
\caption{$M_R = 8$ and $M_T = 64$ SNR -15 dB different resolution of the ADC $b$.}
\label{fig:DLSNR-15EE}
\end{figure}
\begin{figure}
\begin{center}
\begin{tikzpicture}
    \begin{axis}[width=0.95*8.8cm, height=0.95*8.8cm, xlabel={avg. achievable rate [bps/Hz]}, grid, xmin = 1, xmax = 16.2, ymin=10, ymax =115,
                  		ylabel={energy efficiency [bps/J]}, legend pos=north west, legend cell align=left]
		\addplot[ thick, red, mark=*] table[x = y, y=x]{AchievableRateComparison/PlotData/DL/EE/Multipath_8_64_16_32_3.5e-01_1000_SNR_0_EE_DBF.txt};
		\addplot[ thick, green, mark=triangle*] table[x = y, y=x]{AchievableRateComparison/PlotData/DL/EE/Multipath_8_64_16_32_3.5e-01_1000_MAC_2_SNR_0_EE_HBF.txt};
		\addplot[ thick, green, mark=square*] table[x = y, y=x]{AchievableRateComparison/PlotData/DL/EE/Multipath_8_64_16_32_3.5e-01_1000_MAC_4_SNR_0_EE_HBF.txt};
		\addplot[ thick, green, mark=diamond*] table[x = y, y=x]{AchievableRateComparison/PlotData/DL/EE/Multipath_8_64_16_32_3.5e-01_1000_MAC_8_SNR_0_EE_HBF.txt};
		\addlegendentry{DBF}
		\addlegendentry{HBF $M_{RFE}~4$}
		\addlegendentry{HBF $M_{RFE}~2$}
		\addlegendentry{HBF $M_{RFE}~1$}
		\node[anchor=west]  at (axis cs:2.5,15){increasing resolution};
   		\node (source) at (axis cs:2.5,17){};  		
   		\node (destination) at (axis cs:5,35){};
   		\draw[->, very thick](source)--(destination);
     \end{axis}
\end{tikzpicture}
\end{center}
\caption{$M_R = 8$ and $M_T = 64$ SNR 0 dB different resolution of the ADC $b$.}
\label{fig:DLSNR0EE}
\end{figure}
\fi
In this subsection, a downlink like scenario is evaluated. A basestation with 64 antennas ($M_T = 64$) is transmitting to a mobilestation with 8 antennas ($M_R = 8$).
For the channel model the following parameters are used: $L = 32$, $P = 16$, $\beta = 0.35$. For the hybrid beamforming system $M_C \in \{2, 4, 8\}$ and 
therefore $M_{RFE} \in \{4, 2, 1\}$ is used.

\ifCLASSOPTIONdraftcls
Figure \ref{fig:DLULMACSE} (\textbf{A})
\else
Figure \ref{fig:DLMAC4SE}
\fi
shows the average achievable rate for the case of $M_C = 4$ and ADC resolution $b \in \{1, \cdots, 8\}$. The rate curves of the systems including an ADC clearly
converge to the ones assuming no quantization, for higher resolution in both cases of hybrid and digital beamforming. Especially in the low SNR regime (below 0 dB), the
performance of the digital beamforming systems with low resolution ADC (1-3 bit) are very close to the performance without quantization. These systems 
clearly outperform a hybrid beamforming system in this SNR regime. In this evaluation a 4-bit ADC is enough to outperform the hybrid system over the whole SNR range.

Since these system have a different power consumption, we also have to compare the results in terms of energy efficiency. Here we define the energy efficiency (EE) as the 
average achievable rate $R$ divided by the power consumption of the RF-front-end $P_R$:
\begin{equation}
\text{EE} = \frac{R}{P_R}.
\end{equation}
\ifCLASSOPTIONdraftcls
Figure \ref{fig:DLEE} (\textbf{A}) and (\textbf{B}) 
\else
Figure \ref{fig:DLSNR-15EE} and \ref{fig:DLSNR0EE}
\fi
show the energy efficiency over the achievable rate for $M_C \in \{2, 4, 8\}$ and the resolution of the ADC $b \in \{1, \cdots, 8\}$ with SNR $\in$ \{-15 dB, 0 dB\}. 
For both cases, the digital beamforming achieves a higher data rate and also a higher energy efficiency. In the -15 dB SNR case, the difference in energy efficiency is not
substantial but in the 0 dB SNR there is a large gap between hybrid and digital beamforming. In the lower SNR case, the energy efficiency peaks at 3-bit ADC resolution. 
The higher the SNR gets, the larger the ADC resolution that maximizes the energy efficiency. These results show that even when perfect hybrid beamforming without the 
beam-alignment overhead is considered a digital beamforming system is more energy efficient. 
\subsection{Uplink (UL) Point to Point Scenario}
\ifCLASSOPTIONdraftcls
\else
\begin{figure}
\begin{center}
\begin{tikzpicture}
    \begin{axis}[width=0.95*8.8cm, height=0.95*8.8cm, ylabel={avg. achievable rate [bps/Hz]}, grid,  legend cell align=left,
                  		xlabel={SNR [dB]}, xmin =-30, xmax=30, ymin =0, ymax = 80, legend pos=north west, legend cell align=left, legend columns=2]
		\addplot[ thick, red, mark=*] table[x = x, y=y]{AchievableRateComparison/PlotData/UL/SE/Multipath_64_8_16_32_3.5e-01_1000_CapCSITAvg.txt};
		\addplot[ thick, green, mark=triangle*, mark options={solid, fill=green}] table[x = x, y=y]{AchievableRateComparison/PlotData/UL/SE/Multipath_64_8_16_32_3.5e-01_1000_MAC_4_CapHBFNQAvgCombined.txt};
		\addplot[ thick, BTORMIX7] table[x = x, y=y]{AchievableRateComparison/PlotData/UL/SE/Multipath_64_8_16_32_3.5e-01_1000_bits_8_CapAQNMExactCSITAvg.txt};
		\addplot[ thick, dashed, BTOGMIX7] table[x = x, y=y]{AchievableRateComparison/PlotData/UL/SE/Multipath_64_8_16_32_3.5e-01_1000_bits_8_MAC_4_CapHBFAQNMExactCSITAvgCombined.txt};
		\addplot[ thick, BTORMIX6] table[x = x, y=y]{AchievableRateComparison/PlotData/UL/SE/Multipath_64_8_16_32_3.5e-01_1000_bits_7_CapAQNMExactCSITAvg.txt};
		\addplot[ thick, dashed, BTOGMIX6] table[x = x, y=y]{AchievableRateComparison/PlotData/UL/SE/Multipath_64_8_16_32_3.5e-01_1000_bits_7_MAC_4_CapHBFAQNMExactCSITAvgCombined.txt};
		\addplot[ thick, BTORMIX5] table[x = x, y=y]{AchievableRateComparison/PlotData/UL/SE/Multipath_64_8_16_32_3.5e-01_1000_bits_6_CapAQNMExactCSITAvg.txt};
		\addplot[ thick, dashed, BTOGMIX5] table[x = x, y=y]{AchievableRateComparison/PlotData/UL/SE/Multipath_64_8_16_32_3.5e-01_1000_bits_6_MAC_4_CapHBFAQNMExactCSITAvgCombined.txt};
		\addplot[ thick, BTORMIX4] table[x = x, y=y]{AchievableRateComparison/PlotData/UL/SE/Multipath_64_8_16_32_3.5e-01_1000_bits_5_CapAQNMExactCSITAvg.txt};
		\addplot[ thick, dashed, BTOGMIX4] table[x = x, y=y]{AchievableRateComparison/PlotData/UL/SE/Multipath_64_8_16_32_3.5e-01_1000_bits_5_MAC_4_CapHBFAQNMExactCSITAvgCombined.txt};
		\addplot[ thick, BTORMIX3] table[x = x, y=y]{AchievableRateComparison/PlotData/UL/SE/Multipath_64_8_16_32_3.5e-01_1000_bits_4_CapAQNMExactCSITAvg.txt};
		\addplot[ thick, dashed, BTOGMIX3] table[x = x, y=y]{AchievableRateComparison/PlotData/UL/SE/Multipath_64_8_16_32_3.5e-01_1000_bits_4_MAC_4_CapHBFAQNMExactCSITAvgCombined.txt};
		\addplot[ thick, BTORMIX2] table[x = x, y=y]{AchievableRateComparison/PlotData/UL/SE/Multipath_64_8_16_32_3.5e-01_1000_bits_3_CapAQNMExactCSITAvg.txt};
		\addplot[ thick, dashed, BTOGMIX2] table[x = x, y=y]{AchievableRateComparison/PlotData/UL/SE/Multipath_64_8_16_32_3.5e-01_1000_bits_3_MAC_4_CapHBFAQNMExactCSITAvgCombined.txt};
		\addplot[ thick, BTORMIX1] table[x = x, y=y]{AchievableRateComparison/PlotData/UL/SE/Multipath_64_8_16_32_3.5e-01_1000_bits_2_CapAQNMExactCSITAvg.txt};
		\addplot[ thick, dashed, BTOGMIX1] table[x = x, y=y]{AchievableRateComparison/PlotData/UL/SE/Multipath_64_8_16_32_3.5e-01_1000_bits_2_MAC_4_CapHBFAQNMExactCSITAvgCombined.txt};
		\addplot[ thick, blue] table[x = x, y=y]{AchievableRateComparison/PlotData/UL/SE/Multipath_64_8_16_32_3.5e-01_1000_bits_1_CapAQNMExactCSITAvg.txt};
		\addplot[ thick, dashed, blue] table[x = x, y=y]{AchievableRateComparison/PlotData/UL/SE/Multipath_64_8_16_32_3.5e-01_1000_bits_1_MAC_4_CapHBFAQNMExactCSITAvgCombined.txt};

		\addlegendentry{DBF}
		\addlegendentry{HBF}
		\addlegendentry{DBF 8$b$}
		\addlegendentry{HBF 8$b$}
		\addlegendentry{DBF 7$b$}
		\addlegendentry{HBF 7$b$}
		\addlegendentry{DBF 6$b$}
		\addlegendentry{HBF 6$b$}
		\addlegendentry{DBF 5$b$}
		\addlegendentry{HBF 5$b$}
		\addlegendentry{DBF 4$b$}
		\addlegendentry{HBF 4$b$}
		\addlegendentry{DBF 3$b$}
		\addlegendentry{HBF 3$b$}
		\addlegendentry{DBF 2$b$}
		\addlegendentry{HBF 2$b$}
		\addlegendentry{DBF 1$b$}
		\addlegendentry{HBF 1$b$}
     \end{axis}
\end{tikzpicture}
\end{center}
\caption{$M_R = 64$ and $M_T = 8$ $M_{C} = 4$ different resolution of the ADC $b$.}
\label{fig:ULMAC4SE}
\end{figure}
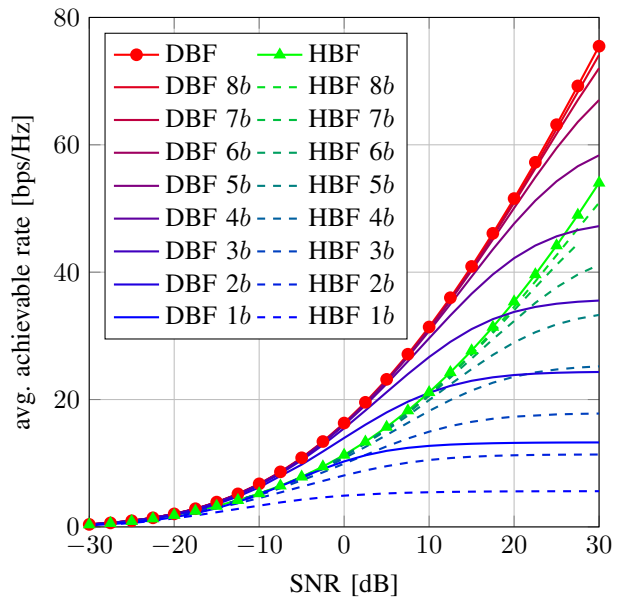
\fi
\ifCLASSOPTIONdraftcls
\begin{figure}
\begin{center}
\begin{tikzpicture}
\begin{groupplot}[group style={group size= 2 by 1, horizontal sep=1.5cm}, width=8cm, height=8cm]
    \nextgroupplot[title={(\textbf{A}) $\text{SNR} = -15\text{dB}$}, xlabel={avg. achievable rate [bps/Hz]}, grid, xmin = 2, xmax = 4, ymin=1.7, ymax =4.5,
                  		ylabel={energy efficiency [bps/J]}, legend pos=south west, legend cell align=left]
		\addplot[ thick, red, mark=*] table[x = y, y=x]{AchievableRateComparison/PlotData/UL/EE/Multipath_64_8_16_32_3.5e-01_1000_SNR_-150_EE_DBF.txt};
		\addplot[ thick, green, mark=triangle*] table[x = y, y=x]{AchievableRateComparison/PlotData/UL/EE/Multipath_64_8_16_32_3.5e-01_1000_MAC_2_SNR_-150_EE_HBF.txt};
		\addplot[ thick, green, mark=square*] table[x = y, y=x]{AchievableRateComparison/PlotData/UL/EE/Multipath_64_8_16_32_3.5e-01_1000_MAC_4_SNR_-150_EE_HBF.txt};
		\addplot[ thick, green, mark=diamond*] table[x = y, y=x]{AchievableRateComparison/PlotData/UL/EE/Multipath_64_8_16_32_3.5e-01_1000_MAC_8_SNR_-150_EE_HBF.txt};
		\node[anchor=west]  at (axis cs:2.15,2.70){increasing resolution};
   		\node (source) at (axis cs:2.25,2.8){};  		
   		\node (destination) at (axis cs:3,3.5){};
   		\draw[->, very thick](source)--(destination);
		\coordinate (c1) at (rel axis cs:0,1);
    \nextgroupplot[title={(\textbf{B}) $\text{SNR} = 0\text{dB}$}, xlabel={avg. achievable rate [bps/Hz]}, grid, xmin = 3, xmax = 16.4, ymin=4, ymax =18,
				legend to name=grouplegend4, legend style={legend columns=4,fill=none,draw=black,anchor=center,align=left, column sep=0.15cm}]
		\addplot[ thick, red, mark=*] table[x = y, y=x]{AchievableRateComparison/PlotData/UL/EE/Multipath_64_8_16_32_3.5e-01_1000_SNR_0_EE_DBF.txt};
		\addplot[ thick, green, mark=triangle*] table[x = y, y=x]{AchievableRateComparison/PlotData/UL/EE/Multipath_64_8_16_32_3.5e-01_1000_MAC_2_SNR_0_EE_HBF.txt};
		\addplot[ thick, green, mark=square*] table[x = y, y=x]{AchievableRateComparison/PlotData/UL/EE/Multipath_64_8_16_32_3.5e-01_1000_MAC_4_SNR_0_EE_HBF.txt};
		\addplot[ thick, green, mark=diamond*] table[x = y, y=x]{AchievableRateComparison/PlotData/UL/EE/Multipath_64_8_16_32_3.5e-01_1000_MAC_8_SNR_0_EE_HBF.txt};
		\addlegendentry{DBF}
		\addlegendentry{HBF $M_{RFE}~32$}
		\addlegendentry{HBF $M_{RFE}~16$}
		\addlegendentry{HBF $M_{RFE}~8$}
		\node[anchor=west]  at (axis cs:4.5,4.7){increasing resolution};
   		\node (source) at (axis cs:5,5){};  		
   		\node (destination) at (axis cs:7.5,8.5){};
   		\draw[->, very thick](source)--(destination);
		\coordinate (c2) at (rel axis cs:1,1);
	\end{groupplot}
    	\coordinate (c3) at ($(c1)!.5!(c2)$);
    	\node[below] at (c3 |- current bounding box.south)
	{\ref{grouplegend4}};
\end{tikzpicture}
\end{center}
\caption{$M_R = 64$ and $M_T = 8$ different resolution of the ADC $b$ and different SNR.}
\label{fig:ULEE}
\end{figure}
\else
\begin{figure}
\begin{center}
\begin{tikzpicture}
    \begin{axis}[width=0.95*8.8cm, height=0.95*8.8cm, xlabel={avg. achievable rate [bps/Hz]}, grid, xmin = 2, xmax = 4, ymin=1.7, ymax =4.5,
                  		ylabel={energy efficiency [bps/J]}, legend pos=south west, legend cell align=left]
		\addplot[ thick, red, mark=*] table[x = y, y=x]{AchievableRateComparison/PlotData/UL/EE/Multipath_64_8_16_32_3.5e-01_1000_SNR_-150_EE_DBF.txt};
		\addplot[ thick, green, mark=triangle*] table[x = y, y=x]{AchievableRateComparison/PlotData/UL/EE/Multipath_64_8_16_32_3.5e-01_1000_MAC_2_SNR_-150_EE_HBF.txt};
		\addplot[ thick, green, mark=square*] table[x = y, y=x]{AchievableRateComparison/PlotData/UL/EE/Multipath_64_8_16_32_3.5e-01_1000_MAC_4_SNR_-150_EE_HBF.txt};
		\addplot[ thick, green, mark=diamond*] table[x = y, y=x]{AchievableRateComparison/PlotData/UL/EE/Multipath_64_8_16_32_3.5e-01_1000_MAC_8_SNR_-150_EE_HBF.txt};
		\addlegendentry{DBF}
		\addlegendentry{HBF $M_{RFE}~32$}
		\addlegendentry{HBF $M_{RFE}~16$}
		\addlegendentry{HBF $M_{RFE}~8$}
		\node[anchor=west]  at (axis cs:2.15,2.70){increasing resolution};
   		\node (source) at (axis cs:2.25,2.8){};  		
   		\node (destination) at (axis cs:3,3.5){};
   		\draw[->, very thick](source)--(destination);
     \end{axis}
\end{tikzpicture}
\end{center}
\caption{$M_R = 64$ and $M_T = 8$ SNR -15 dB different resolution of the ADC $b$.}
\label{fig:ULSNR-15EE}
\end{figure}
\begin{figure}
\begin{center}
\begin{tikzpicture}
    \begin{axis}[width=0.95*8.8cm, height=0.95*8.8cm, xlabel={avg. achievable rate [bps/Hz]}, grid, xmin = 3, xmax = 16.4, ymin=4, ymax =18,
                  		ylabel={energy efficiency [bps/J]}, legend pos=north west, legend cell align=left]
		\addplot[ thick, red, mark=*] table[x = y, y=x]{AchievableRateComparison/PlotData/UL/EE/Multipath_64_8_16_32_3.5e-01_1000_SNR_0_EE_DBF.txt};
		\addplot[ thick, green, mark=triangle*] table[x = y, y=x]{AchievableRateComparison/PlotData/UL/EE/Multipath_64_8_16_32_3.5e-01_1000_MAC_2_SNR_0_EE_HBF.txt};
		\addplot[ thick, green, mark=square*] table[x = y, y=x]{AchievableRateComparison/PlotData/UL/EE/Multipath_64_8_16_32_3.5e-01_1000_MAC_4_SNR_0_EE_HBF.txt};
		\addplot[ thick, green, mark=diamond*] table[x = y, y=x]{AchievableRateComparison/PlotData/UL/EE/Multipath_64_8_16_32_3.5e-01_1000_MAC_8_SNR_0_EE_HBF.txt};
		\addlegendentry{DBF}
		\addlegendentry{HBF $M_{RFE}~32$}
		\addlegendentry{HBF $M_{RFE}~16$}
		\addlegendentry{HBF $M_{RFE}~8$}
		\node[anchor=west]  at (axis cs:4.5,4.7){increasing resolution};
   		\node (source) at (axis cs:5,5){};  		
   		\node (destination) at (axis cs:7.5,8.5){};
   		\draw[->, very thick](source)--(destination);
     \end{axis}
\end{tikzpicture}
\end{center}
\caption{$M_R = 64$ and $M_T = 8$ SNR 0 dB different resolution of the ADC $b$.}
\label{fig:ULSNR0EE}
\end{figure}
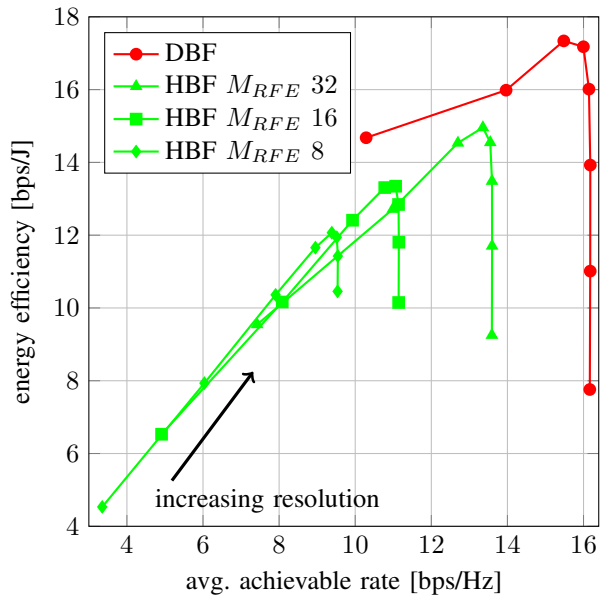
\fi
For the configuration of the system, the same parameters as in the DL like setup in the previous subsection are used. The only difference is that in this 
case the antenna configuration is $M_R = 64$ and $M_T = 8$. 

\ifCLASSOPTIONdraftcls
Figure \ref{fig:DLULMACSE} (\textbf{B})
\else
Figure \ref{fig:ULMAC4SE}
\fi
shows the achievable rate for this case. We observe that the penalty of hybrid beamforming is less severe than in the DL case. 
The reason is that in this case, the side of the system with less antennas (the mobilestation) has no constraints on the front-end which is
the exact opposite in the DL. This means that the number of spatial streams is in all cases just limited by the 8 possible streams of the mobilestation. 
Therefore, the penalty of hybrid beamforming is less and the achievable rate of hybrid and digital beamforming rise with the same slope for 
the case without quantization.

In the low to medium SNR, the low resolution ADC digital beamforming systems perform better than the hybrid beamforming one.
In the high SNR regime, there is no penalty on the number of possible data streams for the hybrid systems, therefore it performs better
in this regime.

\ifCLASSOPTIONdraftcls
Figure \ref{fig:ULEE} (\textbf{A}) and (\textbf{B})
\else
Figure \ref{fig:ULSNR-15EE} and \ref{fig:ULSNR0EE}
\fi
show the energy efficiency over the achievable rate for $M_C \in \{2, 4, 8\}$ and the resolution of the ADC $b \in \{1, \cdots, 8\}$ with SNR $\in$ \{-15 dB, 0 dB\}. 
As in the DL the digital beamforming system is more energy efficient as well as achieving a higher rate. Due to the small number of antennas, the energy efficiency stays 
almost constant for 1 to 3 bit ADC resolution. This can be explained with the fact that if the resolution is small the power consumption of the front-end is dominated by
the other components, and the fact that we have a large degree of freedom with 64 Antennas and therefore the influence of the quantization noise at each of the 
antennas is not very significant. 

%% file: conclusion/conclusion.tex
\section{Conclusion}
The evaluation in this paper showed that low resolution ADC digital beamforming systems are more energy efficient and achieving a higher rate than hybrid
 beamforming systems for the given scenarios, especially in the low to medium SNR region.
We also showed that if the imperfections of the AGC is in the range of -20\% to 20\%, there is no major influence on the
performance. The evaluation of including the off diagonal elements in the quantization error model showed that this could have a
substantial impact on the performance with very low resolution ADCs.

Future extensions should consider the following points. 
For the hybrid beamforming case, the evaluation only shows the result if the beams are already aligned. As shown in \cite{CellSearchDirectionalmmW},
this can be considered to be a large overhead. A possible future mobile broadband system operating at mmWave frequencies will definitely suffer from additional other 
RF-frontend related constraints. Especially considering the inefficiency of the PA that have to operate close to the saturation and therefore introduce 
 distortion to the signal. Also phase noise scales approximately with carrier frequency squared and thus, has to be considered for mmWave systems. 
This will lead to a limited constellation size, which will then bound the overall spectral efficiency. In this evaluation, we also ignored the necessary 
reference overhead for channel estimation. Especially for a high order spatial multiplexing, this is not negligible and will essentially limit the number of spatial data streams.
The channel model is assuming an omnidirectional minimum scattering antenna. Including this consideration into the evaluation would lead to a result that is more close to a practical evaluation. A dynamic multi-user environment would also provide for an interesting comparison between hybrid beamforming and low resolution ADC digital beamforming. 
\section*{Acknowledgment}
This work was supported by the European Commission in the framework of the H2020-ICT-2014-2 project Flex5Gware (Grant agreement no. 671563). 

%% file: appendix/appendix.tex

%


%
%

%% file: literature/literature.tex

\bibliographystyle{IEEEtran}
\bibliography{./literature/IEEEabrv,../../../../bibliography/bibKilian}

%% file: biography/biography.tex
%
%
%
%
%


